\documentclass[a4paper, 11pt]{article}

\usepackage{amssymb,amsthm}
\usepackage{dsfont}
\usepackage[british]{babel} 
\usepackage{verbatim}
\usepackage{graphicx}     
\usepackage{xcolor}
\usepackage{authblk}
\usepackage{tikz}
\usetikzlibrary{positioning,
				decorations.shapes,
				decorations.pathmorphing,
				shapes.geometric,
				shapes.misc,
				backgrounds,
				arrows.meta,
				trees}
\usepackage[shortlabels]{enumitem}
     
\usepackage{booktabs}
\usepackage{caption}
\usepackage{array}
\usepackage{mathtools}
\usepackage{bm}
\usepackage{hyperref}

\definecolor{grigio}{RGB}{80,82,70}
\definecolor{bianco}{RGB}{248,248,240}
\definecolor{mag}{RGB}{249,36,114}
\definecolor{arancio}{RGB}{202,89,36}
\definecolor{ocra}{RGB}{231,219,116}
\definecolor{viola}{RGB}{140,128,255}
\definecolor{verdeacqua}{RGB}{103,216,239}
\definecolor{smeraldo}{RGB}{66,170,150}
\definecolor{matlabyellow}{rgb}{0.9290, 0.6940, 0.1250}
\definecolor{matlabblue}{rgb}{0 0.4470 0.7410}
\definecolor{matlabred}{rgb}{0.6350 0.0780 0.1840}
\definecolor{matlabgreen}{rgb}{0.4660 0.6740 0.1880}

\newcommand\Q{\mathcal Q}
\newcommand\R{\mathbb R}

\newcommand\KK{\mathbb K}
\newcommand\de{\mathrm d}
\newcommand\e{\mathrm e}
\newcommand\der[2]{\frac{\de #1}{\de #2}}                 
\newcommand\pd[2]{\frac{\partial #1}{\partial #2}}         
\renewcommand\*{_\ast}

\let\epsilon=\varepsilon

\theoremstyle{remark}\newtheorem{remark}{Remark}

\DeclarePairedDelimiter{\ins}{\lbrace}{\rbrace}                            
\DeclarePairedDelimiter{\expvalue}{\langle}{\rangle}            
\DeclarePairedDelimiterXPP{\normap}[2]{}{\lVert}{\rVert}{_{#2}}{#1}   

\title{From agent-based models to the macroscopic description of fake-news spread: the role of competence in data-driven applications}

\date{}
\author[1]{J. Franceschi \thanks{jonathan.franceschi01@universitadipavia.it }}
\author[2]{L. Pareschi \thanks{lorenzo.pareschi@unife.it}}
\author[1]{M. Zanella \thanks{mattia.zanella@unipv.it}}

\affil[1]{Department of Mathematics \lq\lq F. Casorati\rq\rq, University of 
Pavia, Italy}
\affil[2]{Department of Mathematics and Computer Science, University of Ferrara, Italy}

\begin{document}
\maketitle

\abstract{%
Fake news spreading, with the aim of manipulating individuals' perceptions of facts, is now recognized as a major problem in many democratic societies.  Yet, to date, little has been understood about how fake news spreads on social networks, what the influence of the education level of individuals is, when fake news is effective in influencing public opinion, and what interventions might be successful in mitigating their effect. In this paper, starting from the recently introduced kinetic multi-agent model with competence by the first two authors, we propose to derive reduced-order models through the notion of social closure in the mean-field approximation that has its roots in the classical hydrodynamic closure of kinetic theory. This approach allows to obtain simplified models in which the competence and learning of the agents maintain their role in the dynamics and, at the same time, the structure of such models is more suitable to be interfaced with data-driven applications. Examples of different Twitter-based test cases are described and discussed.}

\noindent\textbf{Keywords:} fake news spreading, learning dynamics, agent-based 
models, kinetic models, social closure, competence, data uncertainty 

\tableofcontents

\section{Introduction}

Since the 2016 U.S. presidential election, and more recently the COVID-19 infodemic, fake news on social networks, intended to manipulate users' perceptions of events, has been recognized as a fundamental problem in open societies. As fake news proliferate, disinformation threatens democracy and efficient governance. In particular, there is empirical evidence 
that fake news spreads significantly \lq\lq faster, deeper, and more 
widely\rq\rq\ than real news^^>\cite{MIT}. 
In the same study, it is also highlighted that the phenomenon is not due to robotic automatisms of news dissemination but to the actions of human beings sharing the news without the ability to identify misinformation. 

It is therefore of fundamental importance the construction of mathematical models capable of describing such scenarios and with a structure simple enough to be interfaced with data available, for example from social networks, but still embedding the specific features related to the ability of individuals in detecting the piece of false information.  

In recent years, compartmental models inspired by epidemiology have been 
used fruitfully to study spreading phenomena of rumors and hoaxes. For instance, following
the pioneering work of Daley \& Kendall^^>\cite{daley1964}, in^^>\cite{Hong2018}
SIR-type models are used in conjunction with dynamical trust 
rates that account for the different spreading rates in a network. Those 
traditional models were elaborated in^^>\cite{Choi2021}, where the authors 
consider also the impact of online groups in feeding the rumor growth once it 
has started.

Alongside these approaches there are more data-driven works. In this field, 
Twitter has been gaining consensus as a powerful source of useful and 
structured information. A recent example in this direction can be found 
in^^>\cite{Murayama}, that focuses on fake news dissemination on the platform 
using a two-phase model, where fake news initially spread as novel news story 
and after a correction time they are paired with a competitive narrative which 
describes the news as fake in the first place.

Twitter data in conjunction with epidemiological models have already been used 
to study the spread of rumors and fake news by several authors^^>\cite{Fang2013,Fang2014,Maleki2021,cossard2020},
where SIS and SEIZ compartmental models were employed to fit the 
data of the evolution of different news. 
Mounting experimental evidence highlights the strong link between digital media literacy and possibility to reliably identify the quality of online information. This connection has been early identified by communication scientists \cite{glister1997} and later confirmed by experimental studies, see e.g. \cite{guess2020,horrigan2019}. 

In \cite{franceschi2022}, starting from an agent-based model for the dissemination 
of fake news in presence of competence, using the tools of kinetic theory,
in the limit of a large number of agents, novel mathematical models were proposed 
and discussed. Previously, kinetic models that include the role of competence or
knowledge had been proposed in^^>\cite{BT2020,pareschi2014, PVZ2017}. 
The behavior of a 
social system composed by a large number of interacting agents has been studied in the
case of opinion formation^^>\cite{APZ2,cristiani2018,during2015,during2022,toscani2018} and more recently epidemiological dynamics^^>\cite{ABBDPTZ21,APZ1, dimarco2021}. We refer to^^>\cite{pareschi2013BOOK} for an 
introduction to the subject.

The compartmental structure of the model for fake-news spreading in presence 
of competence introduced in^^>\cite{franceschi2022} is composed by four groups of individuals: 
the susceptible (S) agents---defined as the ones who are unaware 
of the fake news; the exposed (E) agents---those who know the news but still 
have not decide whether to spread it or not; the infectious (I) agents---who 
actively divulge and finally the skeptical or removed (R) agents---those who 
are aware of the news but choose to not spread it. On a population divided 
among such categories, there is also a social structure based on an additional time
evolving variable that measures the \emph{competence} level of the agents. 
Although the model has shown the capacity to correctly describe the role of
competence in the dynamics of fake-news, its mathematical structure based on kinetic 
partial differential equations is generally too complex to be
interfaced with the available data. 

In an attempt to address this problem, in the
present work by exploiting the knowledge of the equilibrium states of the corresponding 
mean-field model we derived reduced order macroscopic models
based on ordinary differential equations in which, however, the role of competence
continues to be present. The new social models, thanks to their simpler structure, are
more suitable for data-driven applications. 
We emphasize that the methodology here adopted is quite general and that in principle
points the way to introducing additional social characteristics of individuals into tractable 
mathematical models in terms of structural complexity.

The rest of the manuscript is organized according to the following sections. In Section \ref{sec:model}, we recall the basic concepts of the kinetic model for describing the spread of fake-news in the presence of competence. Next, in Section \ref{sec:red}, using the local equilibrium states of the competence we derive reduced order models that depend on the specific shape of the interaction function. Section \ref{sec:ex} is devoted to presenting a series of numerical experiments in which we first validate the model and then consider data-driven applications based on Twitter. In the last section, a series of final considerations are reported.

\section{Kinetic models, competence and fake news spreading}
\label{sec:model}

In this section we present a model for the description of the spreading of fake news in a society characterized by a heterogeneous competence of agents. Our starting point is the compartmental kinetic approach recently proposed in^^>\cite{franceschi2022}. We suppose that the system of agents can be divided in the following epidemiologically relevant states: susceptible (S) agents are the ones that are unaware of fake news, we further denote as exposed (E) the agents that encountered the fake news but have still to spread them, infectious (I) agents are the real spreader and, finally, the removed (R) agents are not actively engaged in the spread of misinformation. In the following we indicate with $\mathcal C=\{S,E,I,R\}$ the set of epidemiological compartments. 

Aiming to incorporate the effects of personal competence on the fake news dynamics, we stick to a simple mathematical setting where the state of the individuals in each compartment, at any time $t\ge 0$, is characterized by the sole competence level $x\in \mathbb R_+$. Hence, we denote by \mbox{$f_S = f_S(x, t)$}, \mbox{$f_E = 
f_E(x,t)$}, \mbox{$f_I = f_I(x,t)$}, and \mbox{$f_R = f_R(x,t)$} the distribution of  competence at time^^>\mbox{$t \ge0$} of susceptible, exposed, infectious and 
removed individuals, respectively. We neglect natality and 
mortality dynamics since we can consider a short time dynamic where nobody enters or leaves it during the spreading of the fake news. This assumption can be justified based on the average lifespan of fake news. Therefore, we can fix the total distribution of competence of a society to be a probability density for all $t \ge 0$ 
\[
\int_{\mathbb R_+}\left(f_S(x,t) + f_E(x,t) + f_I(x,t) + f_R(x,t) \right)\, \de x = 1,\quad 
t > 0,
\]
Consequently, the quantities
\begin{align*}
S(t) &= \int_{\mathbb R_+} f_S(x,t)\, \de x, &
E(t) &= \int_{\mathbb R_+} f_E(x,t)\, \de x,\\
I(t) &= \int_{\mathbb R_+} f_I(x,t)\, \de x, &
R(t) &= \int_{\mathbb R_+} f_R(x,t)\, \de x
\end{align*}
denote the fractions of the population that are susceptible, exposed, infected, or 
recovered respectively at time $t\ge 0$.  We also denote with $m_J^p(t)$ the moment of the distribution $f_J(x,t)$, $J \in \mathcal C$, of order $p\ge 0$
\begin{align*}
m^p_S(t) &= \frac1{S(t)}\int_{\mathbb R_+} x^p f_S(x,t)\, \de x, &
m^p_E(t) &= \frac1{E(t)}\int_{\mathbb R_+} x^p f_E(x,t)\, \de x, \\
m^p_I(t) &= \frac1{I(t)}\int_{\mathbb R_+} x^p f_I(x,t)\, \de x, &
m^p_R(t) &= \frac1{R(t)}\int_{\mathbb R_+} x^p f_R(x,t)\, \de x.
\end{align*}
Unambiguously we will indicate with $m_J(t)$, $J\in \mathcal C$, the mean values corresponding to $p = 1$. 

\subsection{Competence and learning in multi-agent systems}
Drawing inspiration from seminal models for multi-agent systems in presence of personal competence \cite{pareschi2014,PVZ2017} we introduce a binary interaction term expressing two different processes:
\begin{enumerate}[\textit{\roman*})]
\item \label{item:learnless} learning processes by less competent agents that 
can learn from the more competent ones
\item the competence evolution depends by a social background in which 
individuals grow. 
\end{enumerate}
The dynamics described at point^^>\ref{item:learnless} can be easily sketched 
by the following process: if two agents belonging to compartment $H, J \in 
\mathcal C$ and characterized by competence levels $x,x_* \in \mathbb R_+$ 
meet, their post-interaction competence is given by 
\begin{equation}
\begin{aligned}
\left\lbrace
\begin{aligned}
x'   &= (1 - \lambda_H(x))x + \lambda_{CJ}(x)x\* + \eta_{HJ} x\\
x\*' &= (1 - \lambda_J(x\*))x\* + \lambda_{CH}(x\*)x + \eta_{JH} x\*,
\end{aligned}
\right.
     && H, J \in \mathcal C
     \end{aligned}
\label{eq:cb}
\end{equation}
where $\lambda_H(\cdot)$, $H \in \mathcal C$, quantify the 
amount of competence lost by individuals of compartment^^>$H$ by the natural 
process of forgetfulness and the 
parameter^^>$\lambda_{CH}$, $H \in \mathcal C$, 
models the competence gained through the interaction with members of the 
class^^>$J$, with \mbox{$J \in \mathcal C$}. A possible choice 
for^^>$\lambda_{CJ}(x)$ is^^>\mbox{$\lambda_{CJ}(x) = \lambda_{CJ}\chi(x \ge 
\bar x)$}, where $\chi(\cdot)$ is the characteristic function and \mbox{$\bar x 
\in X$} a minimum level of competence required to the agents for increasing 
their own skills by interactions. In \eqref{eq:cb} $\eta_{HJ}$ and $\eta_{JH}$ 
are centered iid random variable such that, denoting by $\left \langle \cdot 
\right\rangle$ their expectation, we have $\left \langle \eta_{HJ}^2 
\right\rangle = \left \langle \eta_{JH}^2 \right\rangle =  \sigma^2_{HJ}$.

We suppose that the process defined in $(ii)$ takes place in a different time 
scale from the one of interactions between agents.  In particular, unlike 
\cite{franceschi2022} we assume that the time scale of online interactions for 
competence formation is faster than interactions with the social background. To 
this end, we will consider advection terms that will be defined in the next 
section. 

\begin{remark}
It is reasonable to assume that both the processes of gain and loss of 
competence from the interaction with other agents  
in^^>\eqref{eq:cb} are bounded by 
zero. Therefore we suppose that if \mbox{$J, H \in \ins{S, E, I, R}$}, and if 
\mbox{$\lambda_J \in [\lambda_J^-, \lambda_J^+]$}, with \mbox{$\lambda_J^- > 
0$}
and \mbox{$\lambda_J^+ < 1$}, and $\lambda_{CJ}(x) \in [0, 1]$ then 
$\eta_{HJ}$ may, for example, be uniformly distributed 
in^^>\mbox{$[-1 + \lambda_J^+, 1 - \lambda_J^+]$}.
\end{remark}

\subsection{Fake news spreading in presence of a social feature}
\label{subsec:kinetic}
\tikzset{hex/.style={regular polygon,regular polygon sides=6, 
double=grigio, 
rounded corners, draw=bianco, line width = 0.375em,minimum 
size=1.5cm,align=center,inner sep=7.5pt,outer sep=auto}}

Following^^>\cite{franceschi2022} we choose to describe the dissemination of 
fake news through a population of agents via a kinetic compartmental model. In 
this setting the description of the sole spreading dynamics can be illustrated 
by the following system of ODEs
\begin{equation}
\begin{aligned}
\left\lbrace
\begin{aligned}
\frac{dS}{dt} &= -\beta SI + (1 - \alpha)\gamma I\\ 
\frac{dE}{dt} &= \beta SI - \delta E \\
\frac{dI}{dt} &= (1-\eta)\delta E - \gamma I\\
\frac{dR}{dt} &= \eta\delta E + \alpha\gamma I
\end{aligned}
\right.\\[2\smallskipamount]
S + E + I + R = 1.
\end{aligned}\label{eq:seirODE}
\end{equation}
Borrowing from the consolidated epidemiological tradition, we will refer to it 
as the SEIR model. System \eqref{eq:seirODE} describes the evolution of the mass fractions
of the population that belongs to each compartment $J \in \mathcal C$ for each time $t \ge 0$. The parameters appearing in system^^>\eqref{eq:seirODE} are 
presented in Table^^>\ref{tab:parameters}. 
Also a schematic representation of system^^>\eqref{eq:seirODE} is given in 
Figure^^>\ref{fig:seir}. The last equation of system^^>\eqref{eq:seirODE} 
translates the fact that---as specified at the beginning of the Section---the 
total mass of the population is preserved.

The combination of the learning mechanisms presented in the previous subsection 
together with the spreading of the fake news is described by the following 
kinetic model:
\begin{equation}
\left\lbrace
\begin{aligned}
\pd{f_S(x,t)}{t}  &= -K(f_S, f_I)(x,t) + 
(1-\alpha(x))\gamma(x)f_I(x,t)
                     + \dfrac{1}{\epsilon}\sum_{\mathclap{J \in \mathcal C}} 
                     \Q_{SJ}(f_S,f_J)(x,t)\\
                  &\phantom{=} + 
                  \pd{}{x}\left[\left(\lambda_Sx 
                     -\lambda_{BS} m_B\right)f_S(x,t)\right],\\ 
\pd{f_E(x,t)}{t}  &= K(f_S, f_I)(x,t) - \delta(x) f_E(x,t)
                     + \dfrac{1}{\epsilon}\sum_{\mathclap{J \in \mathcal C}}
                     \Q_{EJ}(f_E, f_J)(x,t)\\
                  &\phantom{=} + 
                  \pd{}{x}\left[\left(\lambda_E x
                     -\lambda_{BE} m_B\right)f_E(x,t)\right],\\
\pd{f_I(x,t)}{t}  &= \delta(x)(1 - \eta(x)) f_E(x,t) - \gamma(x)f_I(x,t)
                     +\dfrac{1}{\epsilon} \sum_{\mathclap{J \in \mathcal C}}
                     \Q_{\mathit{IJ}}(f_I, f_J)(x,t)\\
                  &\phantom{=} + 
                  \pd{}{x}\left[\left(\lambda_I x
                     -\lambda_{BI} m_B\right)f_I(x,t)\right],\\
\pd{f_R(x,t)}{t}  &= \delta(x)\eta(x) f_E(x,t) + \alpha(x)\gamma(x) f_I(x,t)
                     + \dfrac{1}{\epsilon}\sum_{\mathclap{J \in \mathcal C}}
                     \Q_{RJ}(f_R, f_J)(x,t)\\
                  &\phantom{=} + 
                  \pd{}{x}\left[\left(\lambda_R x
                     -\lambda_{BR} m_B\right)f_R(x,t)\right],
\end{aligned}
\right.
\label{eq:seisc1}
\end{equation}
where the parameter^^>$\epsilon$ describes the intensity of the 
interactions.
\begin{table}
\centering
\begin{tabular}{cl}
\toprule
 Parameter & Definition\\
\toprule
$\beta$ & contact rate between susceptible and infected individuals\\
$1/\delta$ & average decision time on whether or not to spread fake news\\
$\eta$ & probability of deciding not to spread fake news \\
$1/\gamma$ & average duration of a fake news\\
$\alpha$ & probability of remembering fake news\\
\bottomrule
\end{tabular}
\caption{Parameters definition in the SEIR model \eqref{eq:seisc1}.}
\label{tab:parameters}
\end{table}
In^^>\eqref{eq:seisc1} the functional
\begin{equation}\label{eq:K}
K(f_S, f_I)(x,t) = f_S(x,t) \int_{\R^+} \kappa(x,x\*) 
f_I(x\*,t)\, \de x\*
\end{equation}
is the local incidence rate and $\kappa(x,x\*)$ is a nonnegative contact function measuring the impact of competence in the spreading of fake news. This function is decreasing with respect to the competences $x,x_*\ge 0$ of the population of susceptible and infected agents. In the following we will investigate the macroscopic effects of the following two  choices of $\kappa(x,x\*)$
\begin{enumerate}[\textit{\Alph*}\/)]
\item \label{item:strong} Strong competence-based contact function 
$\kappa(x,x\*) = \beta/(x\,x_*)$, with $\beta>0$,
\item \label{item:weak} Weak competence-based contact function $ \kappa(x,x\*) 
= \beta e^{-x-x_*}$, $\beta >0$.
\end{enumerate}
The two functions are both decreasing but have strong differences for $x,x_*\ll 
1$. Indeed, since^^>\ref{item:strong} is not limited for small competences it 
enforces the spreading of fake news among less competent agents compared with 
$(B)$. Indeed, the function in^^>\ref{item:weak} is bounded in $\mathbb R_+$. 
We further remark that individuals have the highest rates of contact with 
people belonging to the same social class, and thus with a similar level of 
competence.
\begin{remark}
In^^>\eqref{eq:seisc1} the parameters are considered to be dependent on the 
competence level^^>$x$, in general. This is to reflect the fact that competence 
plays a role in the dissemination of fake news.
\end{remark}
\begin{center}
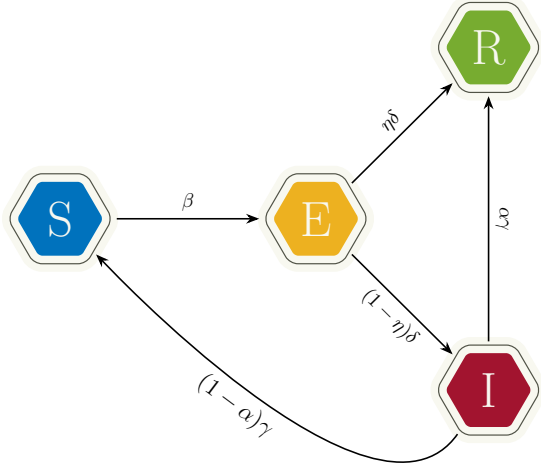
\normalcolor\centering\scalebox{0.75}{%
\begin{tikzpicture}[>=Stealth]
\node[hex,minimum size=1.5cm,fill=matlabblue,text=bianco]
(S) at (0,0) {\Huge S};
\node[hex,minimum size=1.5cm,fill=matlabyellow, right=2.5cm of 
S.east,text=bianco] (E) {\Huge E};
\useasboundingbox (E) circle (0.2\textheight);
\node[hex,minimum size=1.5cm,fill=matlabgreen, above right=2.5cm of 
E.north east,text=bianco] (R) {\Huge R};
\node[hex,minimum size=1.5cm,fill=matlabred, below right=2.5cm of 
E.south east,text=bianco] (I) {\Huge I};
\draw[thick,->] (S) -- (E) node[draw=none,sloped, above, midway]
{$\beta$};
\draw[thick,->] (E) -- (I) node[draw=none,sloped, below, midway]
{$(1 -\eta)\delta$};
\draw[thick,->] (E) -- (R) node[draw=none,sloped, above, midway]
{$\eta\delta$};
\draw[thick,->] (I) -- (R) node[draw=none,sloped, below, midway] 
{\rotatebox{180}{$\alpha\gamma$}};
\draw[thick,->] (I) .. controls +(235:3.5) and +(315:1) .. (S)
node[draw=none, sloped, below, midway] {\large $(1-\alpha)\gamma$};
\end{tikzpicture}}\label{fig:seir}
\captionof{figure}{Dissemination dynamics for the SEIR 
model^^>\eqref{eq:seirODE}}
\end{center}
Furthermore, the operators $\Q_{HJ}(f_H,f_J)(x,t)$, $J\in \mathcal C$,  
describe the binary collisions^^>\eqref{eq:cb} and they determine 
the thermalization of the distribution of competence characterizing the $J$th 
compartment.  The advection terms in^^>\eqref{eq:seisc1} come models the 
influence of the social background on the competence dynamics. It is worth to 
observe that the evolution of mass fractions $J(t)$ 
obeys the classical SEIR model with reinfection^^>\eqref{eq:seirODE} by 
choosing $\Q_{HJ}\equiv 0$ 
and $\kappa(x,x_*) = \beta>0$. This would correspond in considering the 
spreading a fake news independent of the competence level of a system of 
agents. 

In more details, we will consider the operators $\Q_{HJ}$ as integral operators 
that modify the competence distribution through repeated interactions of type 
\eqref{eq:cb} among individuals. We can fruitfully define the introduced 
operators in weak form as follows
\begin{equation}\label{eq:wk1}
\int_{\R^+}\varphi(x) \Q_{HJ}(f_H,f_J) \, \de x = \expvalue*{
                \int_{\R^2_+} \bigl(\varphi(x') - \varphi(x)\bigr)\,f_H(x,t) f_J(x\*,t)
                \, \de x\*\de x
},
\end{equation}
where $\varphi(\cdot)$ is a test function and where the brackets^^>$\expvalue\cdot$  indicate the expectation 
with respect to the random variables^^>$\eta_{HJ},\tilde{\eta}_{HJ}$.  

In the model \eqref{eq:seisc1} the function $\gamma(x)>0$ determines the duration of the fake news and can be strongly influenced by the competence level of the spreader. Furthermore, the function $\delta(x)>0$ is related to the average time that an agent eventually spend before the diffusion of a fake news such that people with high competence invest more time in checking information reliability, and $\eta(x) \in [0,1]$ characterizes individuals' decision to spread fake news. The function $\alpha(x) \in [0,1]$ describes the probability to remember fake news and can be thought less influenced by the competence variable. In Table \ref{tab:parameters} we summarize all the introduced parameters.

\subsection{Asymptotic states of the learning process}
\label{subsec:asymp}
We focus now on the learning dynamics introduced in model \eqref{eq:seisc1} 
whose evolution is given by the nonlinear operators $\Q_{HJ}(f_H,f_J)$, $H,J 
\in \mathcal C$, defined in  \eqref{eq:wk1}. We concentrate in particular on 
the analysis of asymptotic states of the learning dynamics undergoing 
elementary interactions \eqref{eq:cb}. We are therefore interested in the 
asymptotic distribution of the Boltzmann-type model 
\begin{equation}
\label{eq:Bol}
\begin{split}
\dfrac{d}{dt} \int_{\mathbb R_+} \varphi(x) f_H(x,t)dx &=\sum_{J \in \mathcal C}  
                \int_{\R^2_+} \expvalue*{\varphi(x') - \varphi(x)}\,f_H(x,t) f_J(x\*,t)
                \, \de x\*\de x \\
                &= \dfrac{1}{2} \sum_{J \in \mathcal C}  
                \int_{\R^2_+} \expvalue*{\varphi(x')  + \varphi(x_*^\prime)- \varphi(x)-\varphi(x_*)}\,f_H(x,t) f_J(x\*,t)
                \, \de x\*\de x.  \\
\end{split}
\end{equation}
It is easily observed that if $\varphi(x) = 1$ the mass is conserved in \eqref{eq:Bol} corresponding to the conservation of the total number of agents. If $\varphi(x) = x$ in \eqref{eq:Bol} we obtain the evolution of the average competence in each compartment that is not conserved in time
\[\begin{split}
\dfrac{d}{dt} (H(t)m_{H}(t)) &=  \sum_{J \in \mathcal C} \int_{\mathbb R_+^2} \left\langle x^\prime-x\right\rangle f_H(x,t)f_J(x_*,t)\de x\,\de x_* \\
&=H(t) \sum_{J \in \mathcal C} J(t)(\lambda_{CJ}m_J(t) - \lambda_H m_H(t)),
\end{split}\]
and the total competence is conserved
\[
\dfrac{d}{dt} \sum_{H \in \mathcal C} \int_{\mathbb R_+} x f_H(x,t)\de x = 0. 
\]

Since the steady state solution of \eqref{eq:Bol} is difficult to obtain, we can formally derive a simplified Fokker-Planck model in which the study of the asymptotic properties is much easier. To this end, we introduce the following quasi-invariant scaling of the relevant parameter of the binary scheme \eqref{eq:cb} given by
\begin{equation}
\label{eq:qi}
\lambda_H \rightarrow \tau\lambda_H, \qquad \lambda_{CH} \rightarrow \tau\lambda_{CH}, \qquad \sigma_{HJ}^2 \rightarrow \tau \sigma_{HJ}^2, 
\end{equation}
with $\tau>0$. It is worth to mention that the introduced scaling is inspired by the so-called grazing collision limit of the Boltzmann equation, see \cite{Cer,Vil}. In the context of multi-agent systems this scaling has been introduced in \cite{cordier2005JSP,toscani2006CMS}. 

In the introduced regime of parameters the interactions become quasi-invariant, in the sense that the post-interaction competences $(x^\prime,x_*^\prime)$ are such that $x^\prime-x$ and $x_*^\prime-x_*$ are small for $\tau\ll 1$. Hence, assuming $\varphi \in \mathcal C_0$, we can perform the following Taylor expansion
\[
\begin{split}
\varphi(x^\prime)-\varphi(x)  &= (x^\prime-x)\dfrac{\de}{\de x}\varphi(x) + 
\dfrac{1}{2} (x^\prime-x)^2 \dfrac{\de^2}{\de x^2}\varphi(x)+ \dfrac{1}{6} 
(x^\prime-x)^3\dfrac{\de ^3}{\de x^2}\varphi(\bar x), 
\end{split}
\]
with $\bar x \in (\min\{x,x^\prime\},\max\{x,x^\prime\})$. Hence, in the time scale 
$t/\tau$ we have
\begin{multline*}
\frac{\de}{\de t} \int_{\mathbb R_+} \varphi(x) f_H(x,t)\de x \\
=   \sum_{J \in \mathcal C} \int_{\mathbb R^2_+} (-\lambda_H(x)x + 
\lambda_{CJ}(x)x_*)\frac{\de \varphi(x)}{\de x}f_H(x,t)f_J(x_*,t)\,\de x_*\,\de 
x 
\\
\quad + \sum_{J \in \mathcal C}  \frac{\sigma_{HJ}^2}{2} \int_{\mathbb R_+^2}  
x^2 \frac{\de^2\varphi(x)}{\de x^2} f_H(x,t)f_J(x_*,t)\,\de x_*\,\de x + 
\sum_{J \in 
\mathcal C} R_{\varphi}(f_H,f_J),
\end{multline*}
where we exploited the fact that $\left\langle \eta_{HJ} \right\rangle = 0$ and we have defined the sum  of reminder terms
\[
\begin{split}
R_{\tau}(f_H,f_J) &= \frac{1}{2} \int_{\mathbb R_+^2} \tau(-\lambda_H(x)x + 
\lambda_{CJ}(x)x_*)^2 \frac{\de^2\varphi(x)}{\de x^2} f_H(x,t)f_J(x_*,t)\de x 
\,\de x_* \\
&\phantom{{}={}}+ \frac{1}{6} \int_{\mathbb R_+^2} \frac{\left\langle 
(-\lambda_H(x)x + \lambda_{CJ}(x)x_* + \eta_{HJ}x)^3 \right\rangle}{\tau} 
\frac{\de ^3\varphi(x)}{\de x^3} f_H(x,t) f_J(x_*,t)\de x\,\de x_*
\end{split}
\]
We may observe that, assuming $\left\langle 
|\eta_{HJ}|^3\right\rangle<+\infty$, then we may write $\eta_{HJ} = 
\sqrt{\sigma^2}\tilde{\eta}_{HJ}$, where we introduced the centered random 
variable $\tilde{\eta}_{HJ}$ with unitary variance and such that $\left\langle 
|\tilde{\eta}|^3 \right\rangle<+\infty$. Therefore, $\left\langle 
|\eta_{HJ}|^3  \right\rangle = (\sigma^2)^{3/2} \left\langle 
|\tilde{\eta}_{HJ}|^3 \right\rangle$ and, under the scaling \eqref{eq:qi}, we 
get $\left\langle |\eta_{HJ}|^3  \right\rangle \sim \tau^{3/2}\sigma^{3/2}$ . 
Hence, under the above assumption, proceeding as in \cite{cordier2005JSP} we 
can prove that for $\tau\rightarrow 0^+$ 
\[
|R_\varphi(f_H,f_J)(x,t)| \rightarrow 0. 
\]
Therefore in the new time scale, for $\tau\rightarrow 0^+$ and under the quasi-invariant scaling \eqref{eq:qi}, we can show that the solution of model \eqref{eq:Bol} converges to 
\begin{multline}
\frac{\de}{\de t} \int_{\mathbb R_+} \varphi(x) f_H(x,t)\de x \\
=  \sum_{J \in \mathcal C} \int_{\mathbb R^2_+} (-\lambda_H(x)x + 
\lambda_{CJ}(x)x_*)\frac{\de\varphi(x)}{\de x}f_H(x,t)f_J(x_*,t)\,\de x_*\,\de 
x \\
\quad + \sum_{J \in \mathcal C}  \dfrac{\sigma_{HJ}^2}{2} \int_{\mathbb R_+^2}  
x^2 \frac{\de ^2\varphi(x)}{\de x^2} f_H(x,t)f_J(x_*,t)\,\de x_*\,\de x.
\end{multline}
Integrating back by parts we have obtained 
\begin{equation}
\label{eq:FP}
\partial_t f_H(x,t)  = \partial_x \left[ ( \lambda_H(x)x  - \sum_{J \in \mathcal C}  \lambda_{CJ}(x)J(t)m_J(t)) f_H(x,t) + \dfrac{\sigma^2}{2}\partial_x(x^2 f_H(x,t)) \right], 
\end{equation}
with $\sum_{J \in \mathcal C} \sigma_{HJ}^2 = \sigma^2$, coupled with the following boundary conditions
\[
( \lambda_H(x)x  - \sum_{J \in \mathcal C}  \lambda_{CJ}(x)J(t)m_J(t)) f_H(x,t) + \dfrac{\sigma^2}{2}\partial_x(x^2 f_H(x,t)) \Big|_{x = 0} = 0, 
\] 
and 
\[
 \dfrac{\sigma^2}{2}\partial_x(x^2 f_H(x,t)) \Big|_{x=0} = 0,
\]
for all $H \in \mathcal C$. Assuming then $\lambda_{CJ} = \lambda_H = \lambda$ independent by $x \in \mathbb R_+$ for all $J,H \in \mathcal C$ the steady states 
$f_H^\infty(x)$, $H \in \mathcal C$ are solution of 
\[
\lambda( x  -  m) f_H^\infty(x) + \dfrac{\sigma^2}{2}\partial_x(x^2 f_H^\infty(x)) = 0, 
\]
where $m = \sum_{J \in \mathcal C}  J(t)m_J(t)$ is a conserved quantity as we already observed. 
Hence, we obtain that the large time distribution is an inverse Gamma 
\[
f_H^\infty(x) = H \dfrac{k^\mu}{\Gamma(\mu)} \dfrac{e^{-k/x}}{x^{1+\mu}}, 
\]
where
\[
\mu = 1+\dfrac{2\lambda}{\sigma^2}, \qquad k = (\mu-1)m. 
\]

Now, we highlight that for $t \rightarrow +\infty$ we have $f_E(x,t), f_I(x,t) \rightarrow 0$, which means that 
\[
f^\infty(x) = f_S^\infty(x) + f_R^\infty(x). 
\]
In view of $S^\infty + R^\infty = 1$ we conclude that under the introduced assumptions 
\[
f_S^\infty(x) = S^\infty f^\infty, \qquad f_R^\infty = (1-S^\infty)f^\infty. 
\]

\section{Reduced order models for fake news spread with competence}
\label{sec:red}
 
 Once we have characterized the equilibrium distribution of the transition 
 operators $\Q_{HJ}(\cdot,\cdot)$, with $H,J \in \mathcal C$, we can study the 
 complete system \eqref{eq:seisc1}. The aim of this section is the definition 
 of observable macroscopic equations of the introduced kinetic model.
 
 Integrating both sides of  \eqref{eq:seisc1}  with respect to $x \in \mathbb R_+$ and recalling that the introduced operators are mass and momentum preserving, we obtain the following system for the evolution of the mass fractions $J(t)$, $J \in \mathcal C$
 \begin{equation}
\left\lbrace
\begin{aligned}
\der{S(t)}{t}  &= - \int_{\mathbb R^2_+}\kappa(x,x_*)f_S(x,t)f_I(x_*,t)dx\,dx_*
                           + (1-\alpha )\gamma I(t),\\ 
\der{E(t)}{t}  &= \int_{\mathbb R^2_+}\kappa(x,x_*)f_S(x,t)f_I(x_*,t)dx\,dx_* - \delta E(t),\\
\der{I(t)}{t}  &= \delta (1 - \eta ) E(t) -\gamma I(t),\\
\der{R(t)}{t}  &= \delta \eta  E(t) + \alpha \gamma  I(t),
\end{aligned}
\right.
\label{eq:mass}
\end{equation}
whereas for the momentum we get 
 \begin{equation}
\left\lbrace
\begin{aligned}
\der{(S(t)m_S(t))}{t}  &= - \int_{\mathbb R^2_+}x\kappa(x,x_*)f_S(x,t)f_I(x_*,t)dx\,dx_*
                           + (1-\alpha )\gamma I(t)m_I(t),\\ 
\der{(E(t)m_E(t))}{t}  &= \int_{\mathbb R^2_+}x\kappa(x,x_*)f_S(x,t)f_I(x_*,t)dx\,dx_* - \delta E(t),\\
\der{(I(t)m_I(t))}{t}  &= \delta (1 - \eta ) E(t)m_E(t) -\gamma I(t)m_I(t),\\
\der{(R(t)m_R(t))}{t}  &= \delta \eta  E(t)m_E(t) + \alpha \gamma  I(t)m_I(t).
\end{aligned}
\right.
\label{eq:mom}
\end{equation}
We can observe that the obtained system is not closed since the evolution of mass fractions $J(t)$ and of the momentum depend on the evolution of the distribution functions $f_J(x,t)$. The closure of the obtained system can be obtained by formally resorting to a limit procedure. Indeed, assuming that the time scale involved in the process of competence formation is $\epsilon \ll 1$, we have a fast learning process of the system of agents with respect to the evolution of the spreading of fake news. Therefore, for $\epsilon \ll 1$ the distribution function $f_J(x,t)$ reaches fast the inverse Gamma equilibrium with mass fractions $J(t)$ and local mean values $m_J(t)$.

In the following we obtain two different set of macroscopic equations in relation with the considered contact rate function $\kappa(x,x_*)$. 

\subsection{Social closure with a strong competence-based contact function }
\label{subsec:ob1}

We consider the case $(A)$ introduced in Section \ref{subsec:kinetic} corresponding to a strong competence-based contact function defined by $\kappa(x,x_*) = \frac{\beta}{x x_*}$, $\beta>0$. We have

\begin{equation}
\left\lbrace
\begin{aligned}
\der{S(t)}{t}  &= - \beta H_S(t)S(t)H_I(t)I(t)
                           + (1-\alpha )\gamma I(t),\\ 
\der{E(t)}{t}  &= \beta H_S(t)S(t)H_I(t)I(t) - \delta E(t),\\
\der{I(t)}{t}  &= \delta (1 - \eta ) E(t) -\gamma I(t),\\
\der{R(t)}{t}  &= \delta \eta  E(t) + \alpha \gamma  I(t),
\end{aligned}
\right.
\label{eq:seiscompetenza3-bis}
\end{equation}
where 
\begin{equation}\label{eq:HJ}
H_J(t) =  \int_{\R^+} \frac1x f_J(x,t) \, \de x.
\end{equation}
Therefore, in the limit $\epsilon \rightarrow 0^+$ we can plug $f_J^\infty(x)$ in \eqref{eq:HJ} which becomes
\begin{equation}\label{eq:H(t)}
H_J(t) = \frac{\mu}{\mu - 1}\frac1{m_J(t)},
\end{equation}
thanks to the properties of the inverse Gamma distribution, leading to
\begin{equation}
\left\lbrace
\begin{aligned}
\der{S(t)}{t}  &= - \beta \biggl(\frac{\mu}{\mu - 1}\biggr)^2
                    \frac{S(t)I(t)}{m_S(t)m_I(t)} + (1-\alpha )\gamma I(t),\\ 
\der{E(t)}{t}  &=   \beta \biggl(\frac{\mu}{\mu - 1}\biggr)^2
                    \frac{S(t)I(t)}{m_S(t)m_I(t)} - \delta E(t),\\
\der{I(t)}{t}  &= \delta (1 - \eta ) E(t) -\gamma I(t),\\
\der{R(t)}{t}  &= \delta \eta  E(t) + \alpha \gamma  I(t).
\end{aligned}
\right.
\label{eq:seirAfr}
\end{equation}

Next, looking at^^>\eqref{eq:mom}, recalling that under the 
hypothesis that $\lambda_J = \lambda_{CJ}$ for $J \in \mathcal{C}$, the 
knowledge exchange operator also preserves momentum, we have the following 
system of equations
\begin{equation}
\left\lbrace
\begin{aligned}
\der{[m_S(t)S(t)]}{t}  &= - \beta S(t)\frac{\mu}{\mu - 1}\frac{I(t)}{m_I(t)}
                           + (1-\alpha )\gamma m_I(t)I(t)
                       +\lambda  (m_B - m_S(t)) S(t),\\ 
\der{[m_E(t)E(t)]}{t}  &= \beta S(t)\frac{\mu}{\mu - 1}\frac{I(t)}{m_I(t)} -
                        \delta m_E(t)E(t)
                       +\lambda  (m_B - m_E(t)) E(t),\\
\der{[m_I(t)I(t)]}{t}  &= \delta (1 - \eta ) m_E(t)E(t) -\gamma m_I(t)I(t)
                       +\lambda  (m_B - m_I(t)) I(t),\\
\der{[m_R(t)R(t)]}{t}  &= \delta \eta  m_E(t)E(t) + \alpha \gamma  m_I(t)I(t)
                       +\lambda  (m_B - m_R(t)) R(t),
\end{aligned}
\right.
\label{eq:seircompetenzameansmasses}
\end{equation}
which, using the fact that
\[
\der{[m_J(t)J(t)]}{t} = J(t) \der{m_J(t)}t + m_J(t) \der{J(t)}t,
\]
implies
\begin{equation}
\left\lbrace
\begin{aligned}
\der{m_S(t)}{t}  &=\beta\biggl(\frac{\mu}{\mu-1}\biggr)\frac{I(t)}{m_I(t)}
                    \frac1{\mu - 1} + (1-\alpha )\gamma \frac{I(t)}{S(t)}
                    \biggl[m_I(t) - m_S(t) \biggr]\\
                 &\phantom{{}={}}+\lambda (m_B - m_S(t)),\\ 
\der{m_E(t)}{t}  &=\beta \biggl(\frac{\mu}{\mu - 1}\biggr)
                    \frac{S(t)I(t)}{E(t)m_I(t)}
                    \biggl[1 - \frac{\mu}{\mu - 1}\frac{m_E(t)}{m_S(t)}
                      \biggr]
                    +\lambda (m_B - m_E(t)),\\
\der{m_I(t)}{t}  &= \delta (1 - \eta )\frac{E(t)}{I(t)}(m_E(t) - m_I(t))
                    +\lambda  (m_B - m_I(t)),\\
\der{m_R(t)}{t}  &= \delta \eta  \frac{E(t)}{R(t)}(m_E(t) - m_R(t)) + \alpha 
                    \gamma  \frac{I(t)}{R(t)}(m_I(t) - m_R(t))\\
                 &\phantom{{}={}}+\lambda (m_B - m_R(t)),
\end{aligned}
\right.
\label{eq:seirAme2}
\end{equation}
that is, we obtained a closed system of eight ordinary differential 
equations^^>\eqref{eq:seirAfr}, \eqref{eq:seirAme2}.

\subsection{Social closure with a weak competence-based contact function}

If, instead, we consider the case^^>\ref{item:weak} of 
Section^^>\ref{subsec:kinetic}, corresponding to the weak 
competence-based contact function defined by $\kappa(x,y) = \e^{-x}\e^{-y}$, it is possible to write
\begin{equation}
\label{eq:HJwe}
\tilde H_J(t) = \int_{\mathbb R_+} e^{-x}f_J(x,t)dx. 
\end{equation}
As discussed in Section \ref{subsec:ob1}, in the limit $\epsilon \rightarrow 0^+$ we may plug the asymptotic distribution $f_J^\infty$ of the Fokker-Planck model \eqref{eq:FP} in \eqref{eq:HJwe} to obtain
\[
\begin{split}
\tilde H_J(t) = \underbrace{\frac{2(\mu-1)^{\mu/2} 
                                  (m_J(t))^{\mu/2}\,\KK_\mu(2\sqrt{(\mu-1)m_J(t)})}
                                       {\Gamma(\mu)}}_{\coloneqq C_\mu(m_J)},
\end{split}\]
where $\KK_a(x)$ stands for the modified Bessel function of the second kind of 
order^^>$a$ evaluated at^^>$x$. Hence,  if we consider system^^>\eqref{eq:mass} under the assumption of weak competence-based contact function we obtain
\begin{equation}
\left\lbrace
\begin{aligned}
\der{S(t)}{t}  &= - \beta \tilde H_S(t)S(t)\tilde H_I(t)I(t)
                           + (1-\alpha )\gamma I(t),\\ 
\der{E(t)}{t}  &= \beta \tilde H_S(t)S(t) \tilde H_I(t)I(t) - \delta E(t),\\
\der{I(t)}{t}  &= \delta (1 - \eta ) E(t) -\gamma I(t),\\
\der{R(t)}{t}  &= \delta \eta  E(t) + \alpha \gamma  I(t),
\end{aligned}
\right.
\label{eq:seiscompetenza3-bisexp}
\end{equation}
which becomes
\begin{equation}
\left\lbrace
\begin{aligned}
\der{S(t)}{t}  &= - \beta C_\mu(m_S)\,C_\mu(m_I)S(t)I(t) + (1-\alpha )\gamma 
I(t),\\
\der{E(t)}{t}  &=   \beta C_\mu(m_S)\,C_\mu(m_I)S(t)I(t) - \delta E(t),\\
\der{I(t)}{t}  &= \delta (1 - \eta ) E(t) -\gamma I(t),\\
\der{R(t)}{t}  &= \delta \eta  E(t) + \alpha \gamma  I(t).
\end{aligned}
\right.
\label{eq:seirBfr2}
\end{equation}
The next equation will help to close the system
\begin{equation}\label{eq:integralexexp}
\int_{\R^+} x\,f_J^\infty(t)\e^{-k/x}\e^{-x}\, \de x
      = m_J(t) J(t) C_{\mu - 1}(m_J),
\end{equation}
which is a straightforward consequence of the following property of the 
modified Bessel functions of the second kind
\[
x\, \KK_{\mu+1}(x) - 2\mu \KK_\mu(x) = x\, \KK_{\mu-1}(x), \quad x\in \R^+.
\]
Again under the assumptions that $\lambda_J = \lambda_{CJ}=\lambda$ for $J \in 
\mathcal{C}$, integrating with respect to $x$ equation^^>\eqref{eq:mom}, with 
the aid of equation^^>\eqref{eq:integralexexp}, we get
\begin{equation}
\left\lbrace
\begin{aligned}
\der{[m_S(t)S(t)]}{t}  &= - \beta C_{\mu-1,S}C_{\mu,k}S(t)m_S(t) I(t)
                          + (1-\alpha )\gamma m_I(t)I(t)\\ 
                       &\phantom{{}={}}+\lambda (m_B - m_S(t)) S(t),\\
\der{[m_E(t)E(t)]}{t}  &= \beta C_{\mu-1,S}C_{\mu,k}S(t)m_S(t) I(t)
                          - \delta m_E(t)E(t)\\
                       &\phantom{{}={}} +\lambda  (m_B - m_E(t)) E(t),\\
\der{[m_I(t)I(t)]}{t}  &= \delta (1 - \eta ) m_E(t)E(t) -\gamma m_I(t)I(t)
                         +\lambda  (m_B - m_I(t)) I(t),\\
\der{[m_R(t)R(t)]}{t}  &= \delta \eta  m_E(t)E(t) + \alpha \gamma  m_I(t)I(t)
                         +\lambda  (m_B - m_R(t)) R(t),
\end{aligned}
\right.
\label{eq:seirmeansmassesexp}
\end{equation}
which, using again the fact that
\[
\der{[m_J(t)J(t)]}{t} = J(t) \der{m_J(t)}t + m_J(t) \der{J(t)}t,
\]
leads to
\begin{equation}
\left\lbrace
\begin{aligned}
\der{m_S(t)}{t}  &= -\beta C_{\mu,I}\, m_S(t) I(t)
                    \bigl[C_{\mu - 1, S}\,m_S - C_{\mu,S}\bigr]
                    + (1-\alpha)\gamma\frac{I(t)}%
                    {S(t)}\bigl[m_I(t) - m_S(t)\bigr]\\
                 &\phantom{{}={}} +\lambda  (m_B - m_S(t)),\\ 
\der{m_E(t)}{t}  &= \beta C_{\mu,I}\,\frac{I(t)S(t)}{E(t)}
                     \bigl[C_{\mu-1,S}\, m_S(t) - C_{\mu,S}\,m_E(t)\bigr]
                  + \lambda  (m_B - m_E(t)),\\
\der{m_I(t)}{t}  &= \delta (1 - \eta )\frac{E(t)}{I(t)}(m_E(t) - m_I(t))
                    +\lambda  (m_B - m_I(t)),\\
\der{m_R(t)}{t}  &= \delta \eta  \frac{E(t)}{R(t)}(m_E(t) - m_R(t)) + \alpha 
                    \gamma  \frac{I(t)}{R(t)}(m_I(t) - m_R(t))
                    +\lambda  (m_B - m_R(t)).
\end{aligned}
\right.
\label{eq:seirBme2}
\end{equation}

\section{Examples and applications}
\label{sec:ex}
In this section we numerically validate the modeling framework  proposed in 
\eqref{eq:seisc1} with local incidence rate \ref{eq:K} in the settings 
\ref{item:strong}-\eqref{item:weak}. We stress that those form of contact 
functions generate different macroscopic models that have been defined in 
\eqref{eq:seirAfr},\eqref{eq:seirAme2} and 
\eqref{eq:seirBfr2},\eqref{eq:seirBme2}, respectively, for $\epsilon \ll 1$. 
Once  established the consistency of the approach, we proceed by exploiting the 
macroscopic sets of equations for calibration purposes based on a freely 
available repository for the spreading of hashtags linked to known fake news. 
The proposed data-oriented approach is fundamental to experimentally observe 
the different impact of the contact function in identifying impact of 
competence in the fake news dynamics. 

From the numerical point of view we will exploit an implicit structure 
preserving method for the Fokker-Planck operator \eqref{eq:FP} based on the 
schemes presented in \cite{pareschi2018}. The advantage of these methods relies 
on an arbitrarily accurate description of the steady state distribution of the 
Fokker-Planck model of interest. Similar approaches have been investigated in a 
different context also in \cite{dimarco2021,dimarco2022,pareschi2022}. 

\subsection{Test 1: Validation of the social closure}
In this first test we compare the evolution of mass fractions $J(t)$ and  
means $m_J(t)$, $J \in \mathcal C$, obtained from direct integration of $f_J(x,t)$, solution to \eqref{eq:seisc1}, with respect to the competence $x \in \mathbb R_+$, with the macroscopic models \eqref{eq:seirAfr},\eqref{eq:seirAme2} and \eqref{eq:seirBfr2}--\eqref{eq:seirBme2} for several regimes of $\epsilon>0$. We start by outlining the procedure by which we solve the system of kinetic equations \eqref{eq:seisc1} with Fokker-Planck interaction operators. Since $\epsilon>0$ is assumed to be small, we adopt a time splitting procedure. In particular, upon introducing a time discretization $t^n = n\Delta t$, $\Delta t>0$ constant, we proceed as follows. 

\begin{figure}
\centering
\includegraphics[scale=0.5]{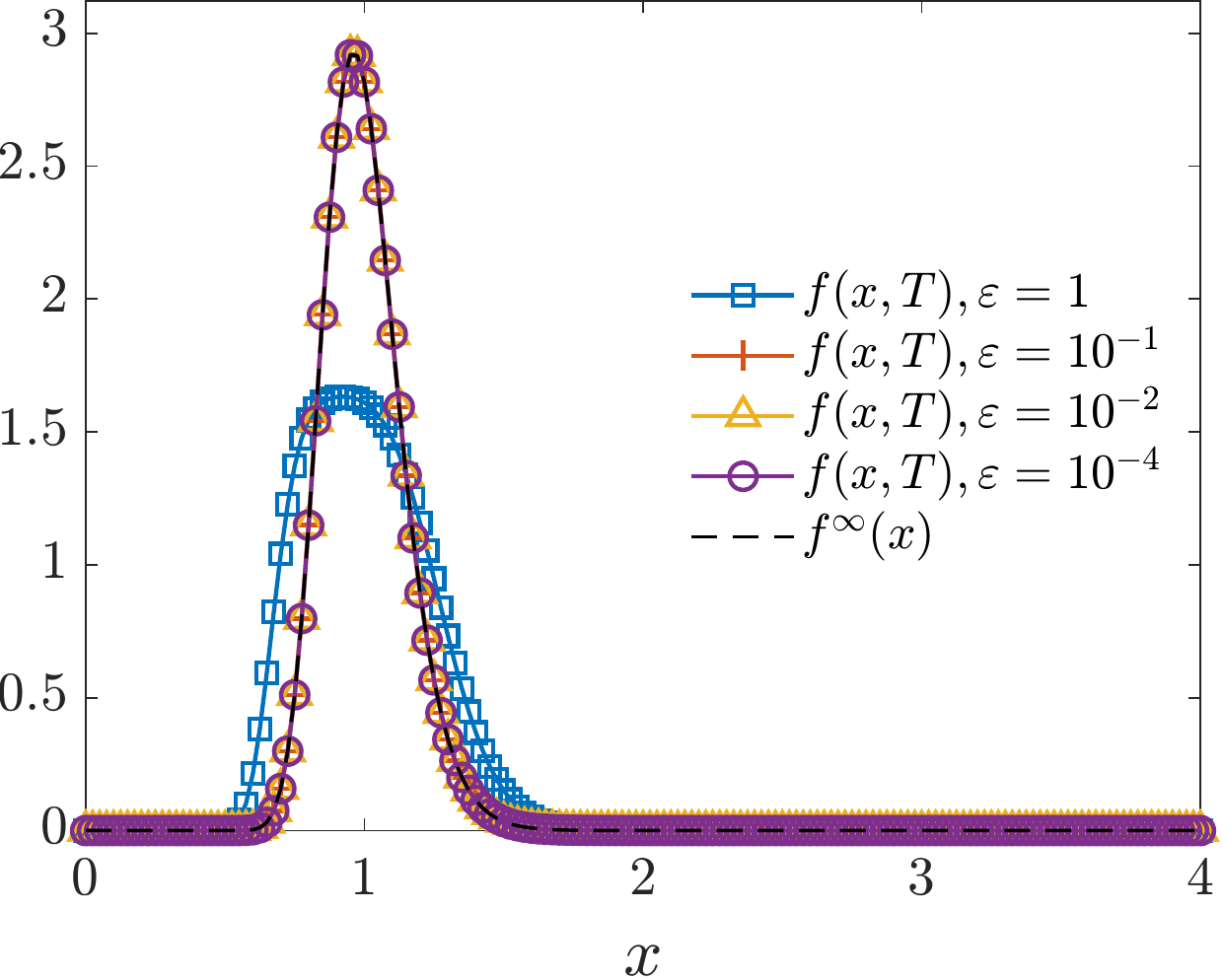}
\includegraphics[scale=0.5]{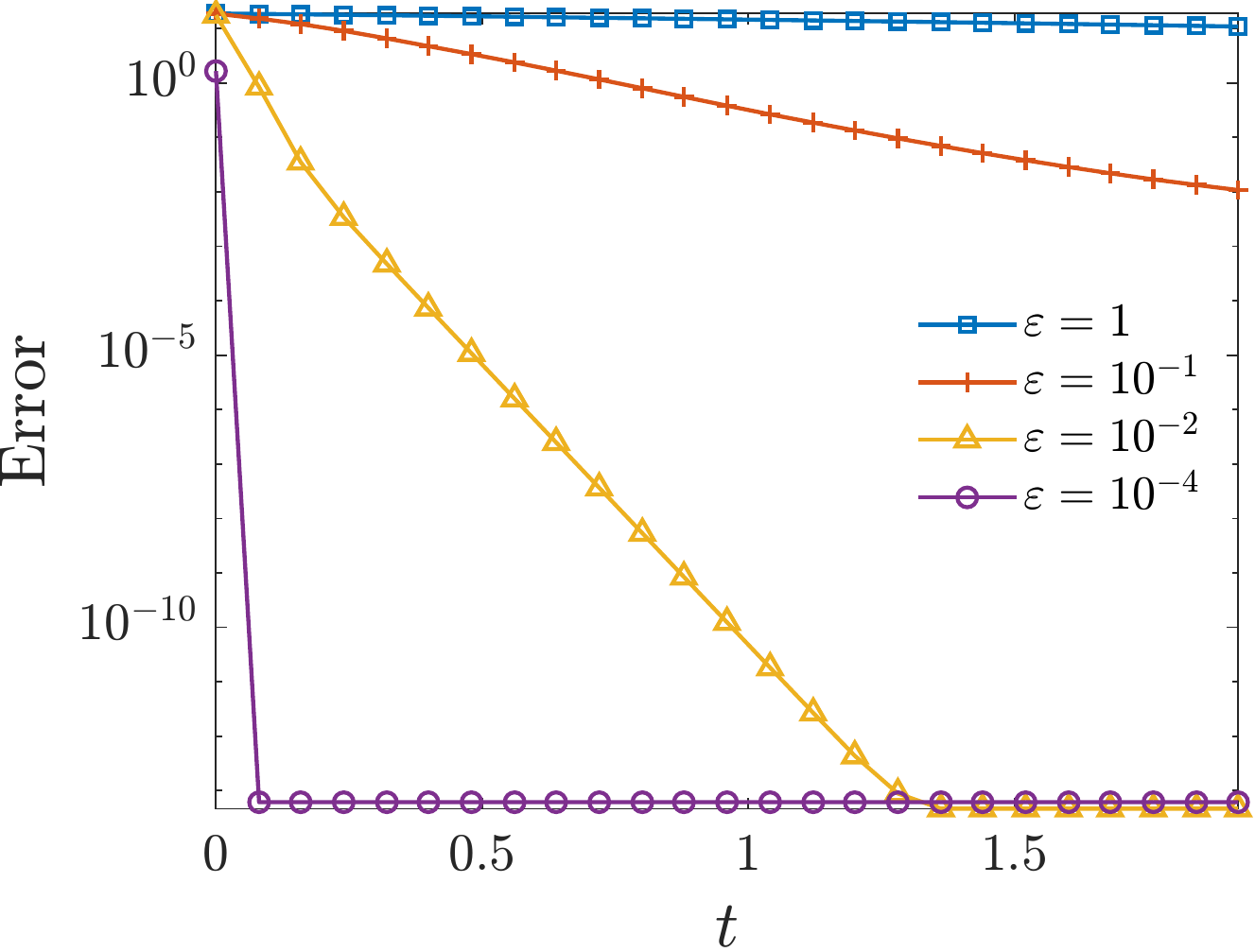}
\caption{\textbf{Test 1}. Left: numerical distribution obtained with SP 
implicit scheme at time $T = 2$ for several $\epsilon = 
1$, $10^{-1}$, $10^{-2}$, $10^{-4}$. Right: evolution of the $L^2$ error 
$\|f(x,T)-f^\infty(x) \|_{L^2}$. In all the tests we considered a 
discretization of the domain $[0,4]$ obtained with $N_v = 201$ grid points and 
$\Delta t = 10^{-4}$. }
\label{fig:steady}
\end{figure}

\paragraph{I. Fokker-Planck solver.}
At time $t = t^n$, we determine the distributions $f_H(x,t)$ for all $H \in \mathcal C$ solution to 
\[
\left\lbrace
\begin{aligned}
\partial_t F_H(x,t) &= \frac{1}{\epsilon} \Q(F_H)(x,t), \quad t \in 
(t^n,t^{n+1/2}] \\
F_H(x,t^n) &= F_H^0(t^n,x), 
\end{aligned}
\right.\label{eq:FP1}
\]
where $\Q(f_H)$ is the Fokker-Planck operator defined in Section 
\ref{subsec:asymp} whose form, in the hypothesis $\lambda_H = \lambda_{CJ} = 
\lambda$,  is given by
\[
\Q(F_H)(x,t) =\partial_x \left[ \lambda(x-m)F_H + \dfrac{\sigma^2}{2}\partial_x 
(x^2F_H(x,t)) \right]. 
\]
In this step we take advantage of an implicit structure preserving (SP) scheme 
for Fokker-Planck equations \cite{pareschi2018} and describes with arbitrary 
accuracy the steady state of the model. In Figure \ref{fig:steady} we report 
for several $\epsilon = 1,10^{-1},10^{-2},10^{-4}$ the numerical solution of 
the Fokker-Planck model in the time interval $[0,T]$, $T = 2$, obtained from a 
discretization of the domain $[0,4]$ with $N_v = 201$ grid points and with 
$\Delta t = 10^{-4}$. We may observe that the scheme is capable to approximate 
the inverse Gamma analytical equilibrium $f^\infty(x)$. We also report the 
evolution of the $L^2$ numerical error computed as $\|f(x,t)-f^\infty(x) 
\|_{L^2}$ in the time frame $[0,2]$ from which we can observe how for 
sufficiently small values of $\epsilon $ we correctly approximate the given 
equilibrium distribution. 

\paragraph{II. Advection-Reaction step.} Hence, we consider the distribution obtained in the interaction step as an input for the advection-reaction dynamics for $t \in [t^{n+1/2},t^{n+1}]$
\[\left\lbrace
\begin{aligned}
\partial_t f_S(x,t)  &= -K(f_S, f_I)(x,t) + 
(1-\alpha)\gamma f_I(x,t)+ 
                \lambda \partial_x \left[\left(x 
                     - m_B\right)f_S(x,t)\right],\\ 
\partial_t f_E(x,t)  &= K(f_S, f_I)(x,t) - \delta f_E(x,t) + 
                 \lambda \partial_x \left[\left( x
                     - m_B\right)f_E(x,t)\right],\\
\partial_t f_I(x,t)  &= \delta(1 - \eta) f_E(x,t) - \gamma f_I(x,t) + 
                 \lambda \partial_x \left[\left( x - m_B\right)f_I(x,t)\right],\\
\partial_t f_R(x,t)  &= \delta\eta f_E(x,t) + \alpha\gamma f_I(x,t)  + \lambda  \partial_x \left[\left( x
                     - m_B\right)f_R(x,t)\right]\\
f_H(x,t^{n+1/2}) &= F_H(x,t^{n+1/2}), \quad H \in \{S,E,I,R\}.
\end{aligned}
\right.
\] 
In particular, we adopted a second order Lax-Wendroff scheme coupled with an explicit time integration. 

\begin{figure}
\centering
\includegraphics[scale = 0.4]{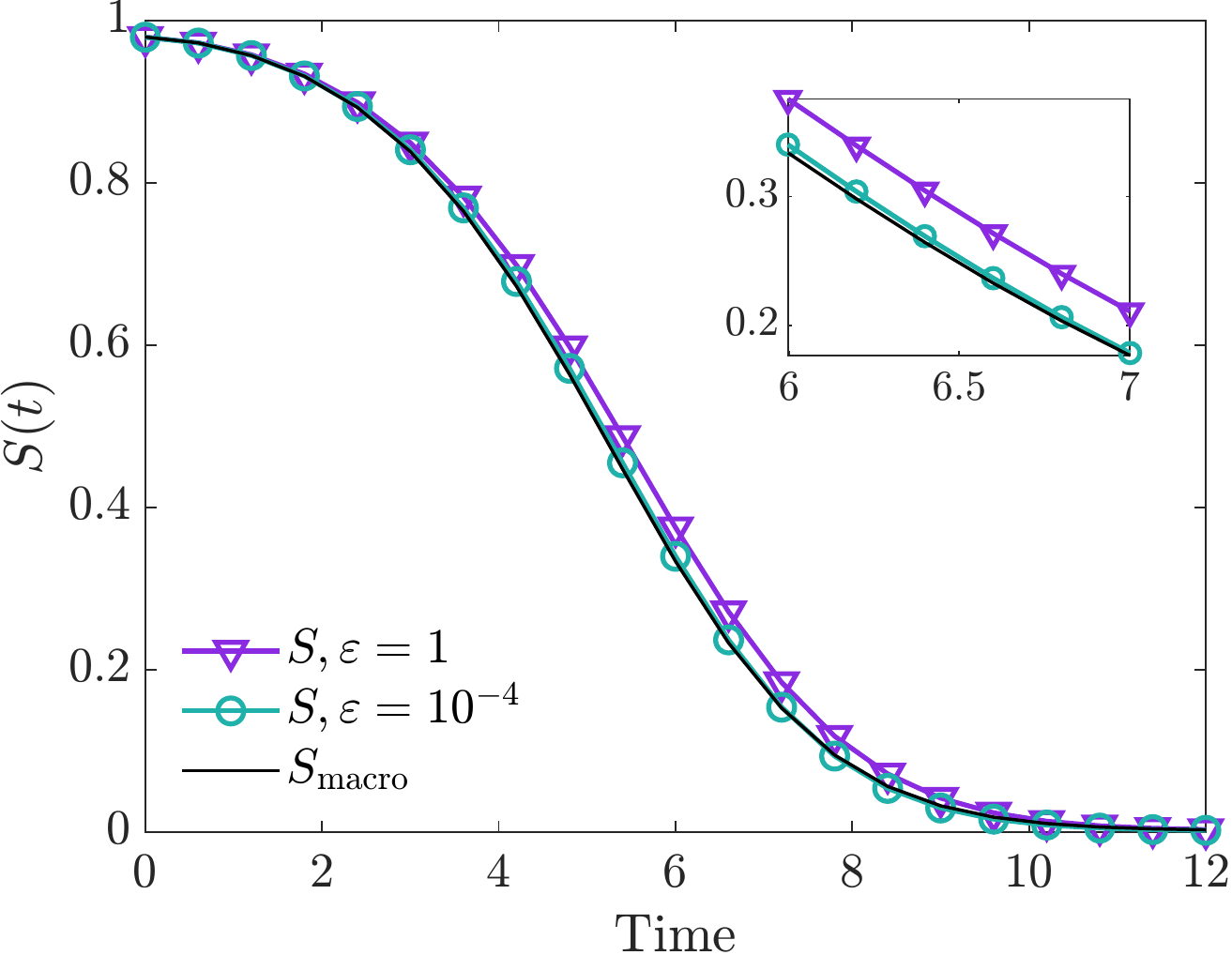}%
\includegraphics[scale = 0.4]{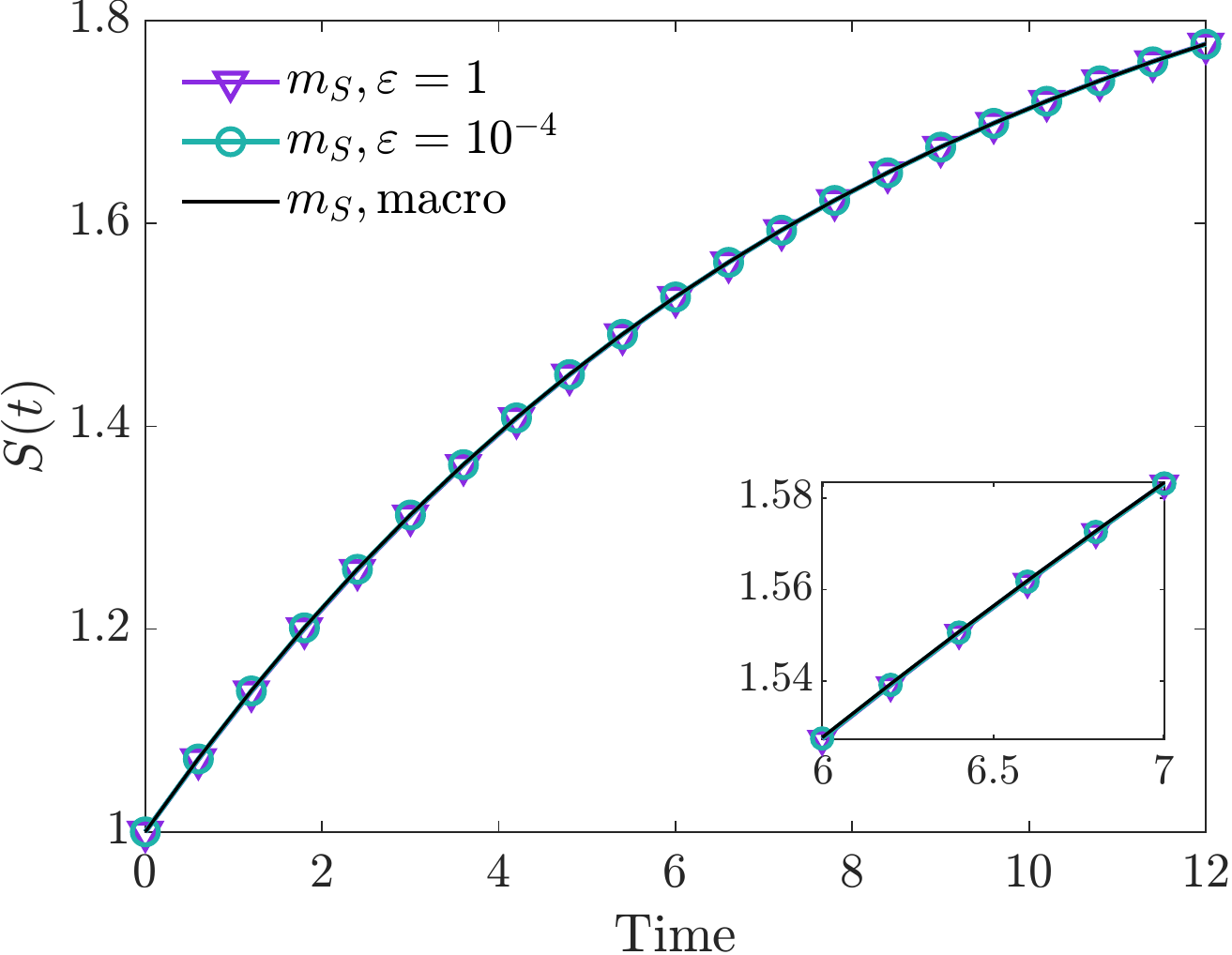}\\
\includegraphics[scale = 0.4]{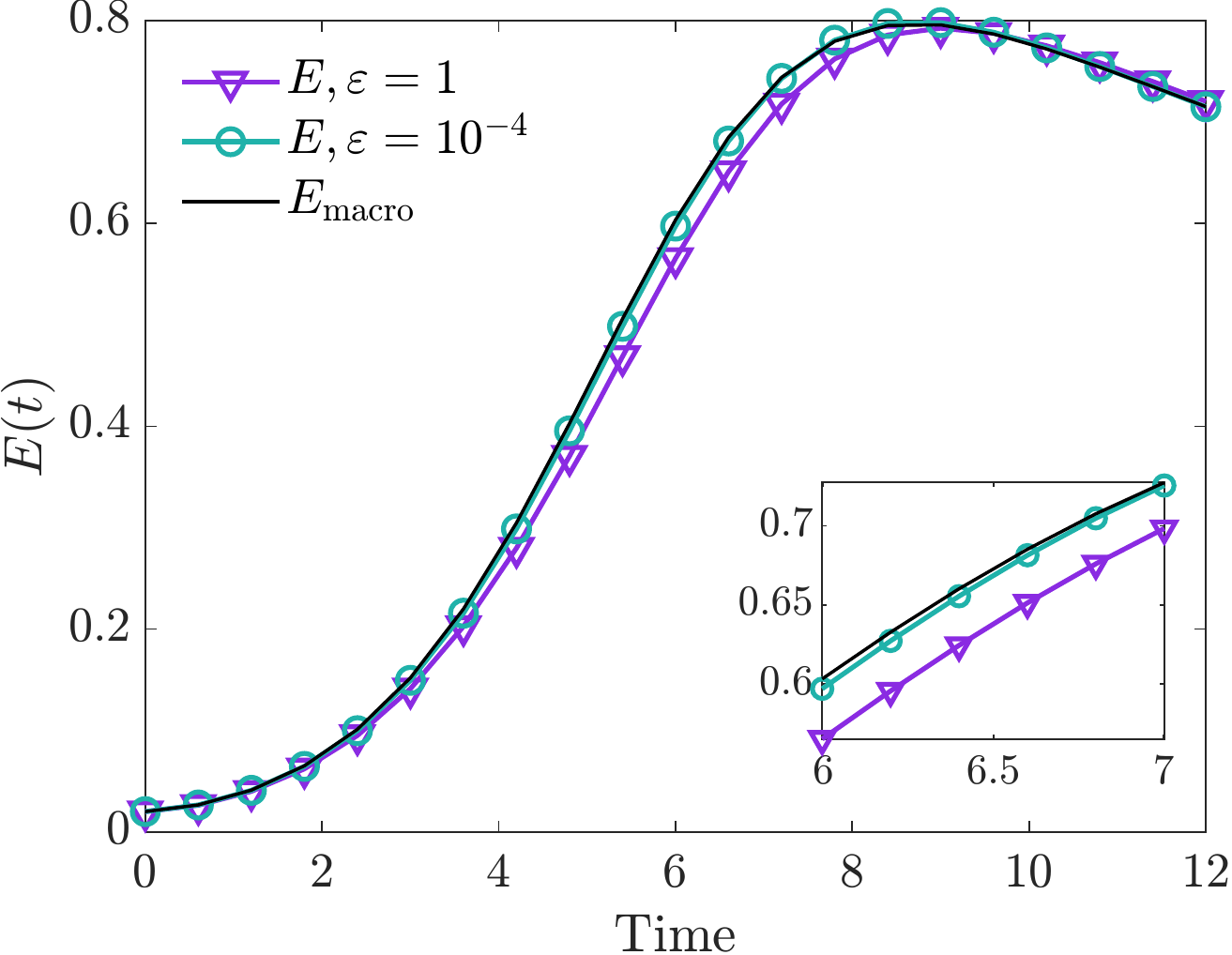}%
\includegraphics[scale = 0.4]{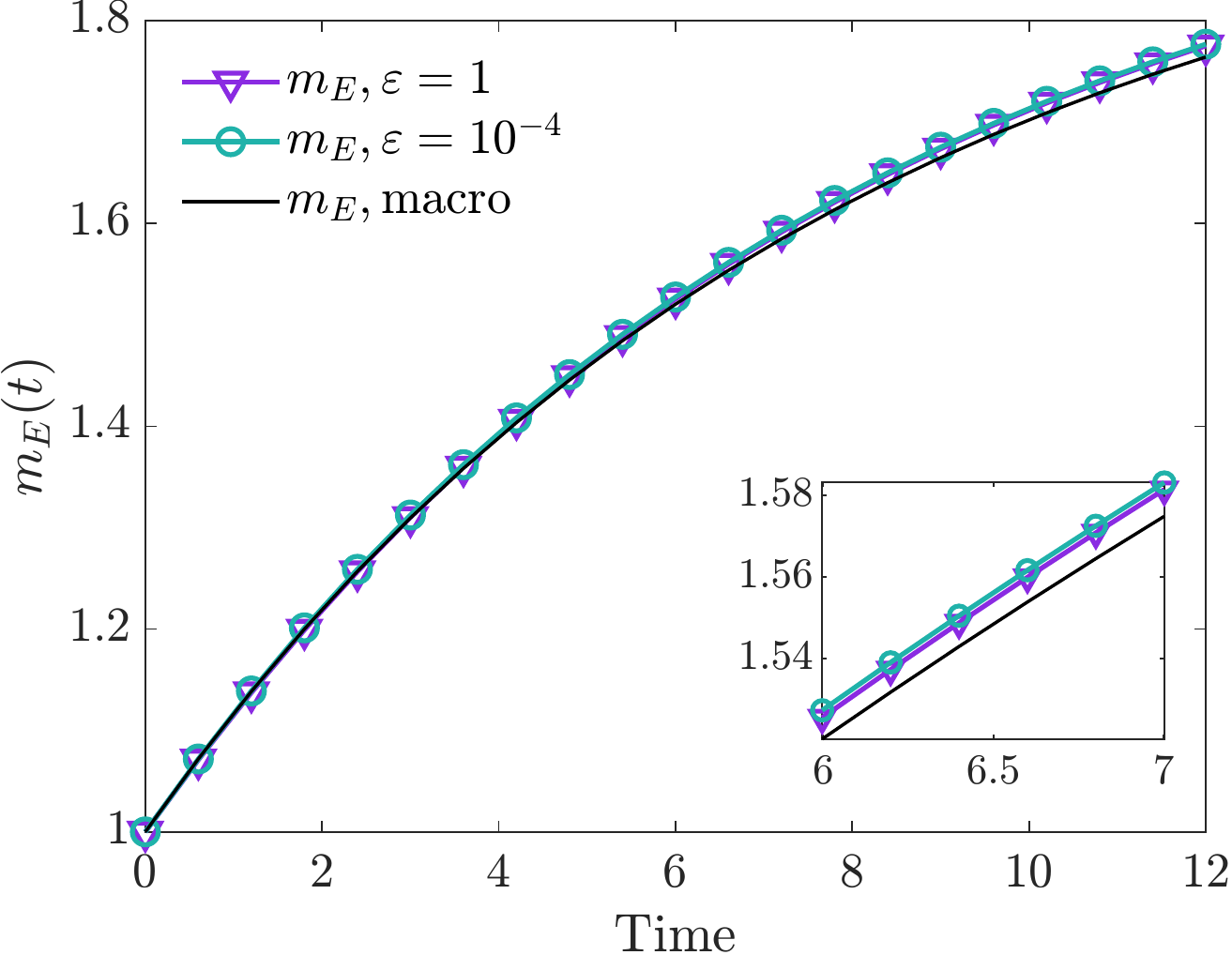}\\
\includegraphics[scale = 0.4]{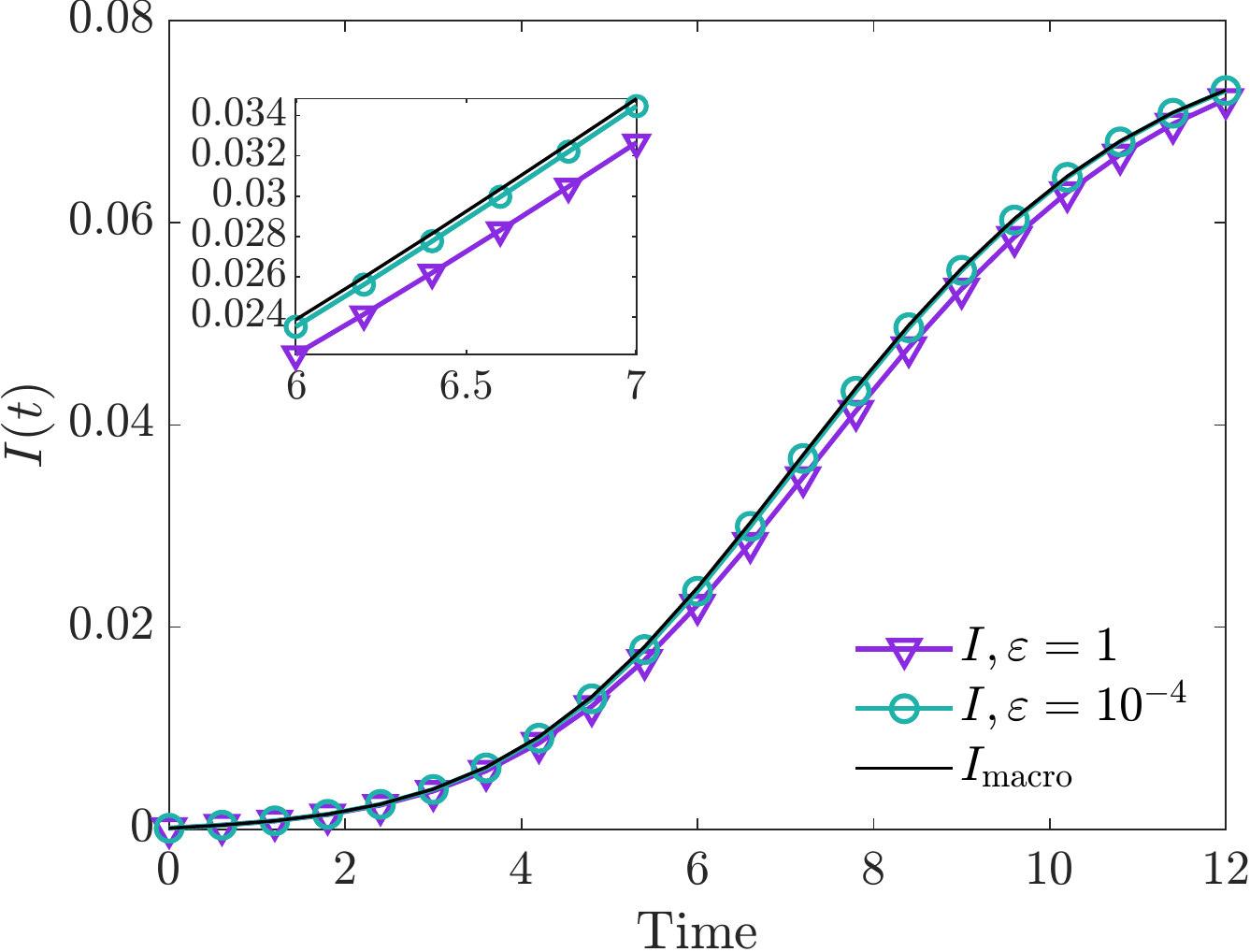}%
\includegraphics[scale = 0.4]{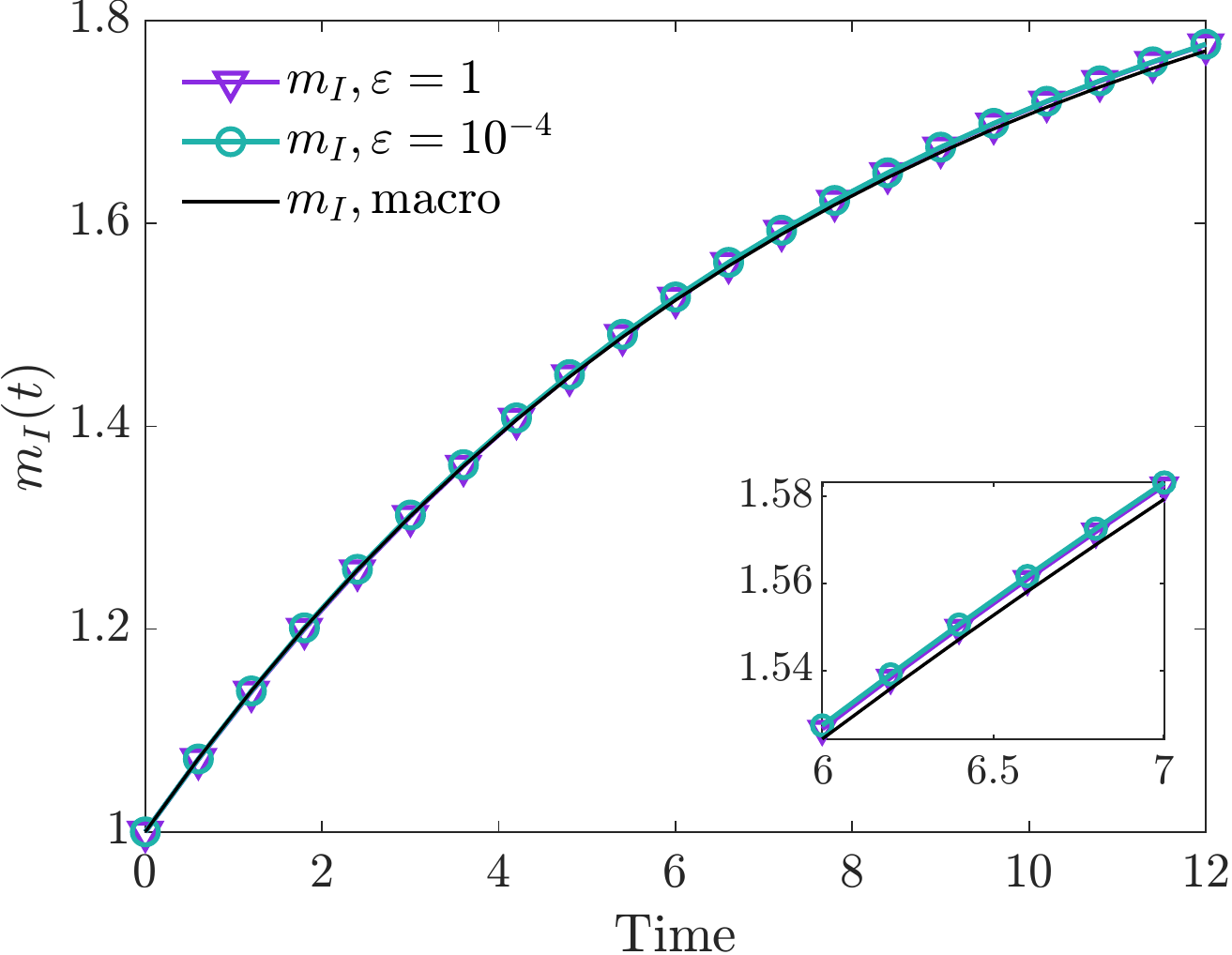}\\
\includegraphics[scale = 0.4]{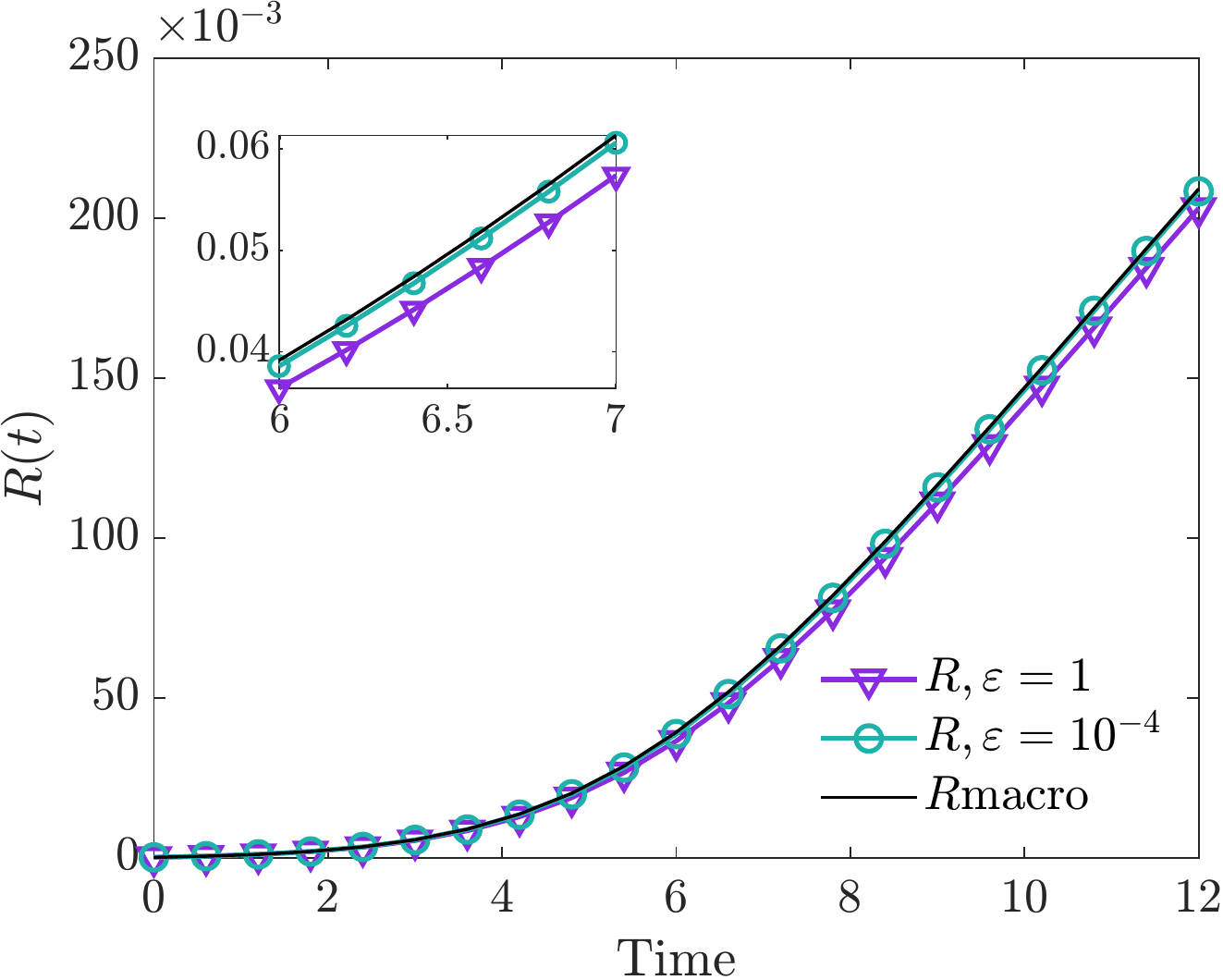}%
\includegraphics[scale = 0.4]{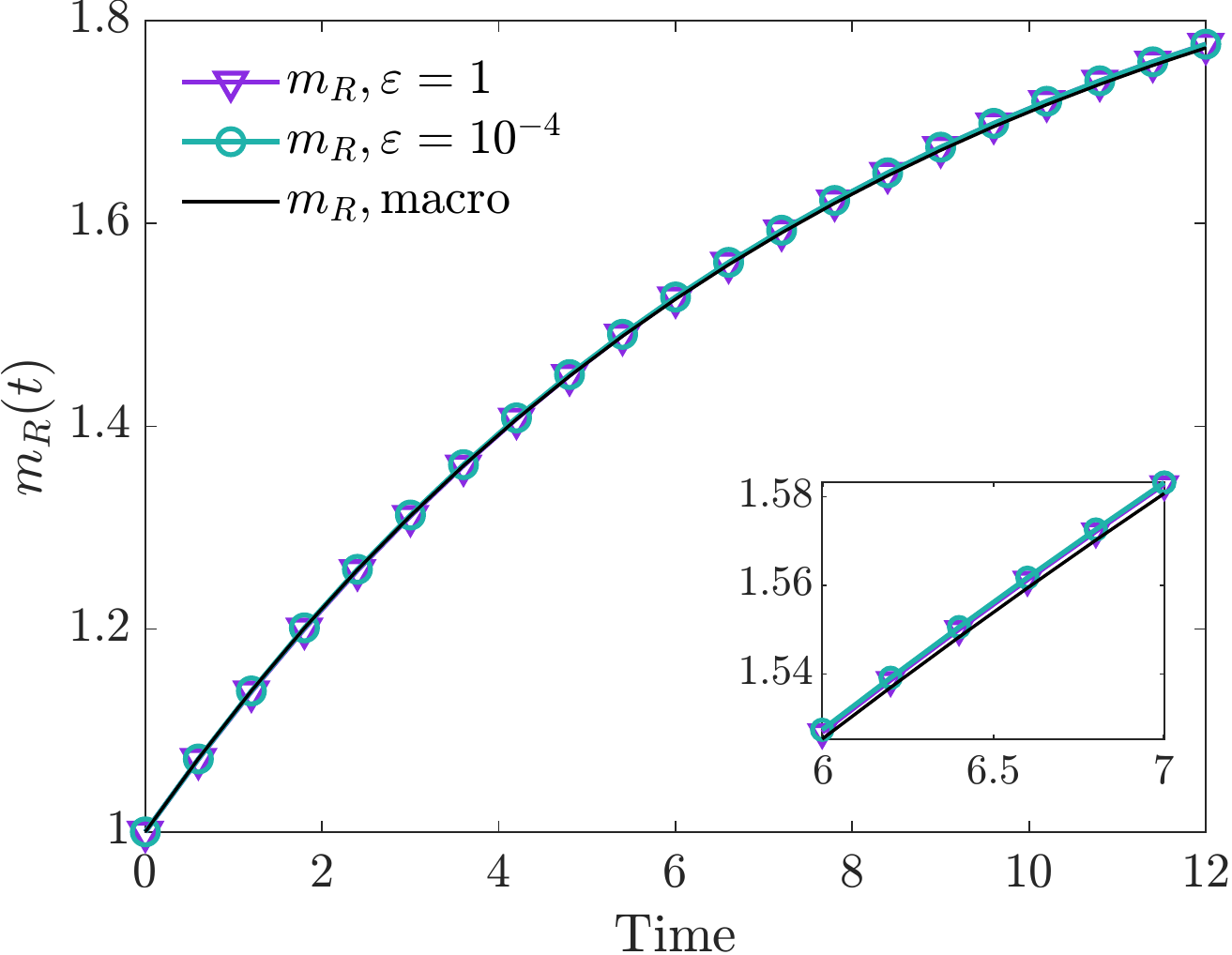}
\caption{\textbf{Test 1}. Evolution of mass fractions (left) and mean values (right) obtained from direct integration of the kinetic model \eqref{eq:seisc1} in the case $\kappa(x,x_*) = 
\beta/(x\,x_*)$, for $\epsilon = 1$, 
$\epsilon = 10^{-4}$ together with the evolution of mass fractions of the 
macroscopic model \eqref{eq:seirAfr}--\eqref{eq:seirAme2}. In both cases we 
considered $\alpha = 0.9$, $\beta = 20$, $\gamma = 0.2$, $\delta = 0.05$. The 
kinetic model has been solved through the scheme $\mathrm I$--$\mathrm{II}$ 
over the domain $[0,4]$, discretization obtained with $N_v = 201$ grid points 
and $\Delta t = 10^{-4}$. The initial distribution of the kinetic model has 
been defined in \eqref{eq:f0_t1}-\eqref{eq:m0_t1}. }
\label{fig:epsilon-ip}
\end{figure}

\begin{figure}
\centering
\includegraphics[scale = 0.4]{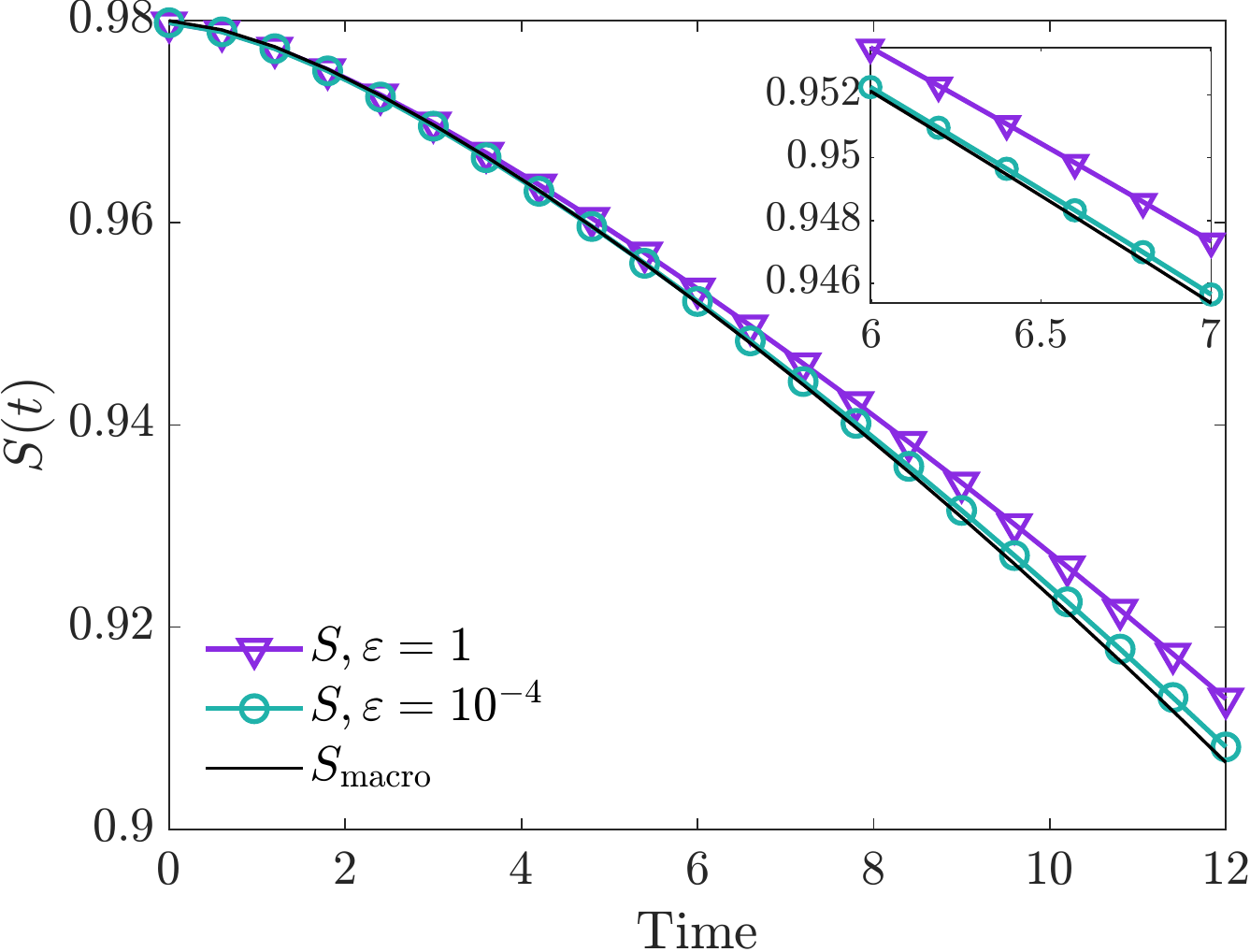}%
\includegraphics[scale = 0.4]{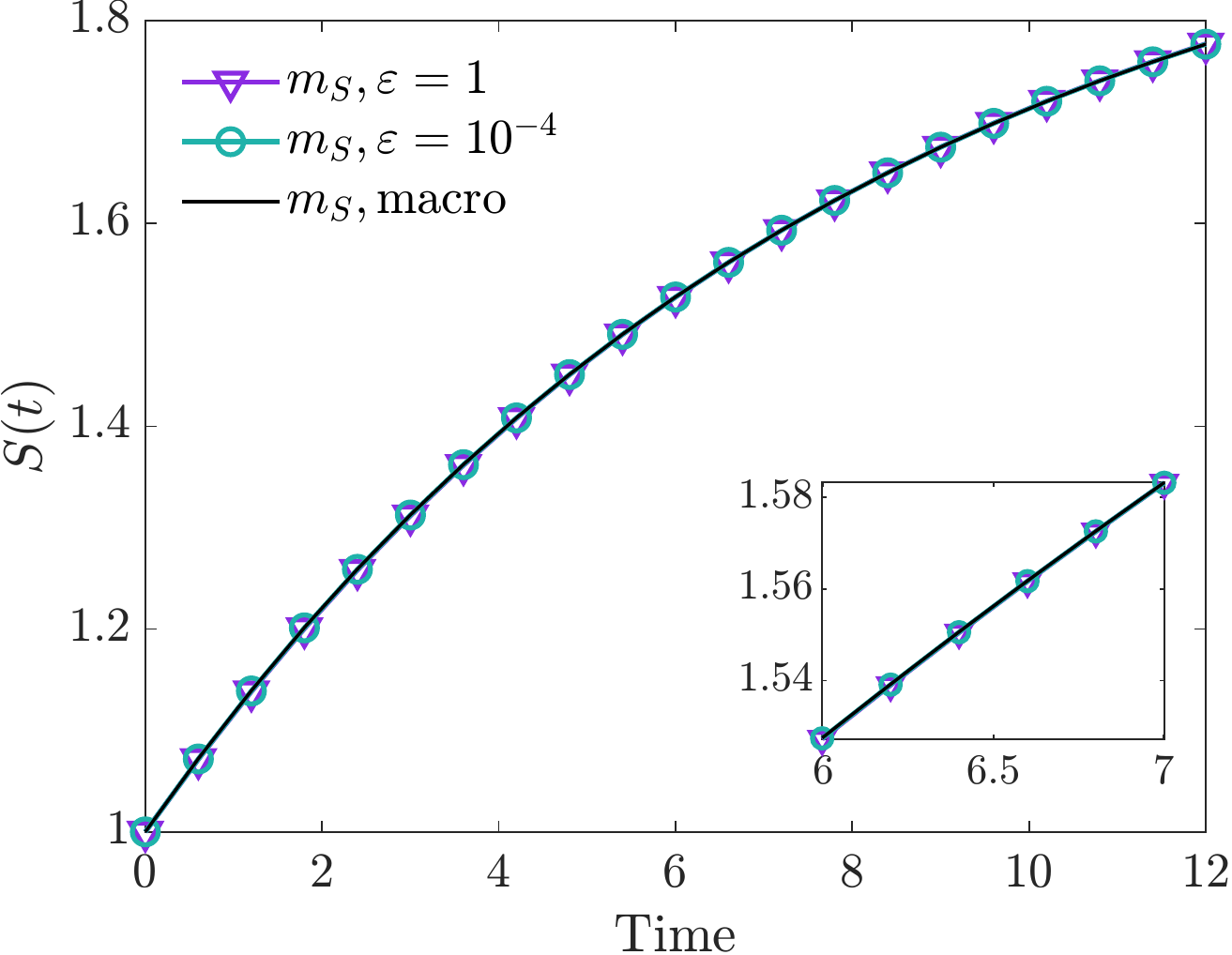}\\
\includegraphics[scale = 0.4]{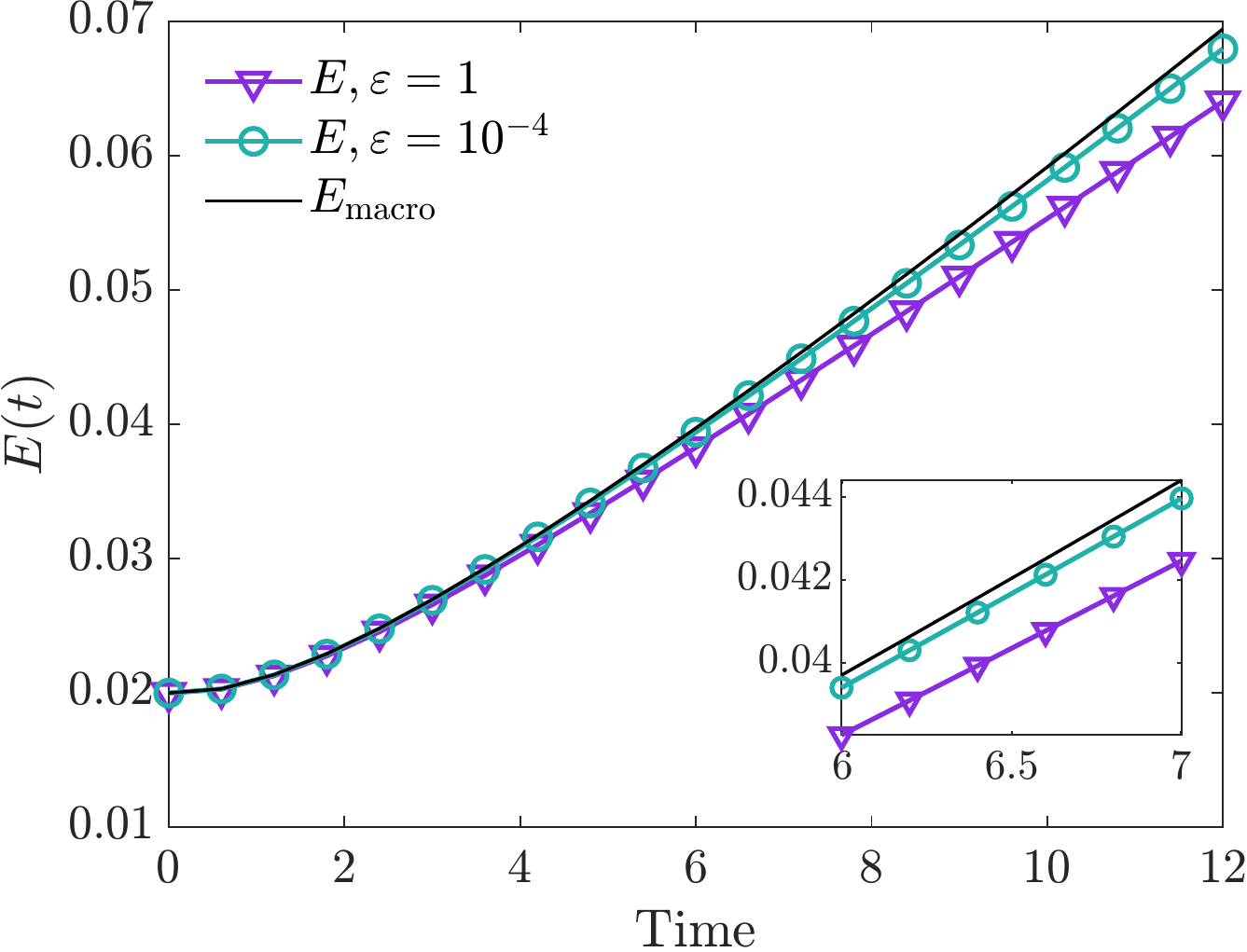}%
\includegraphics[scale = 0.4]{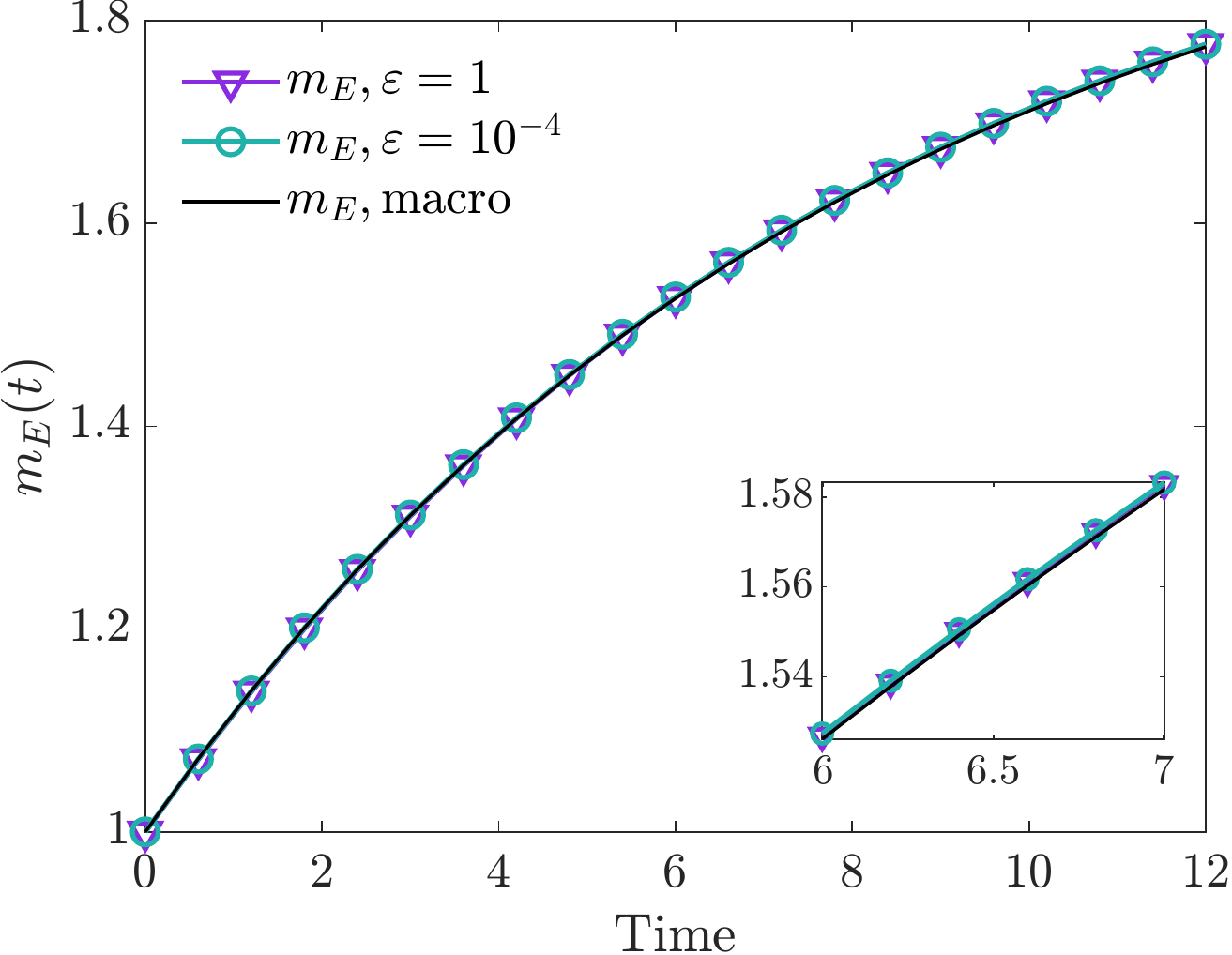}\\
\includegraphics[scale = 0.4]{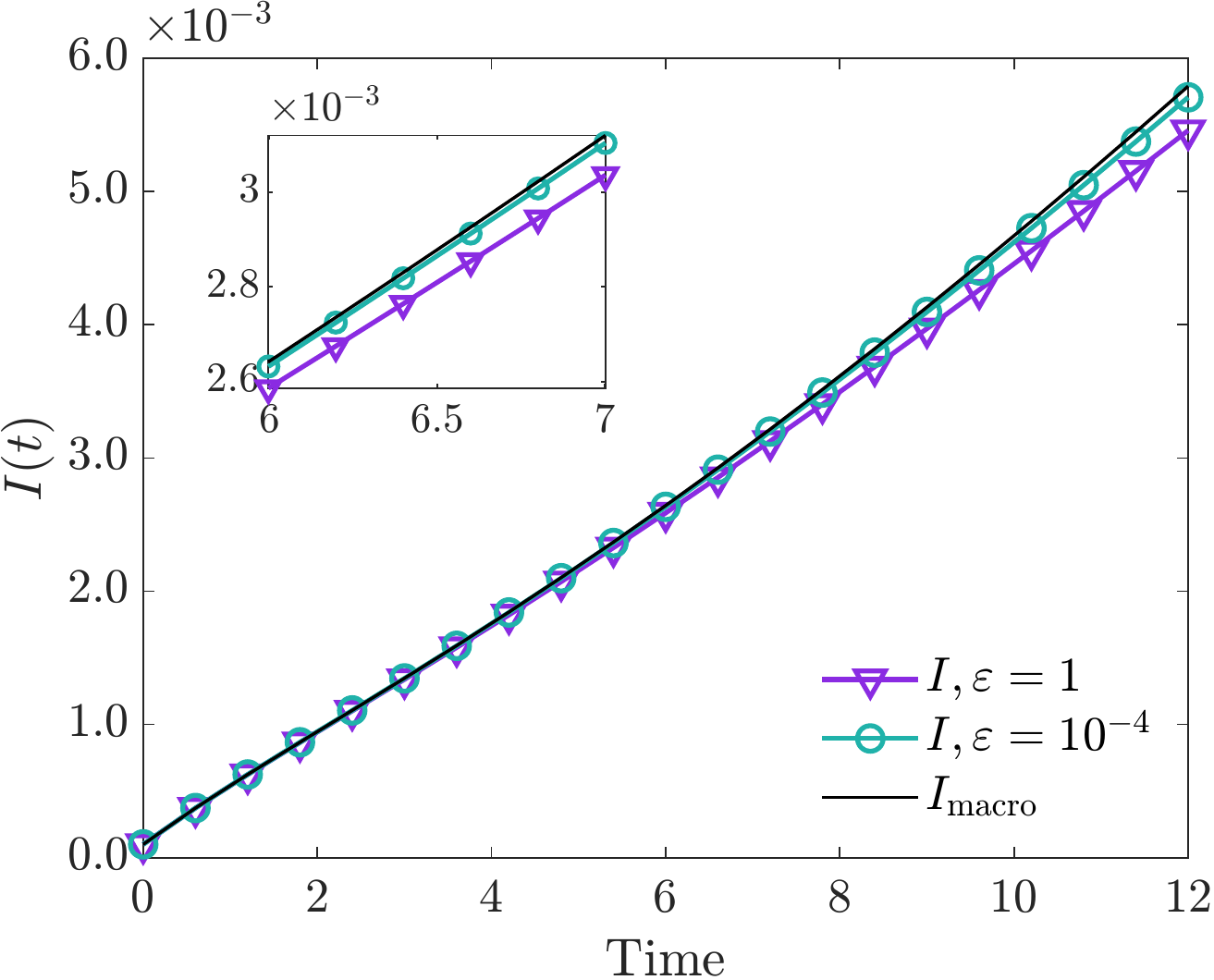}%
\includegraphics[scale = 0.4]{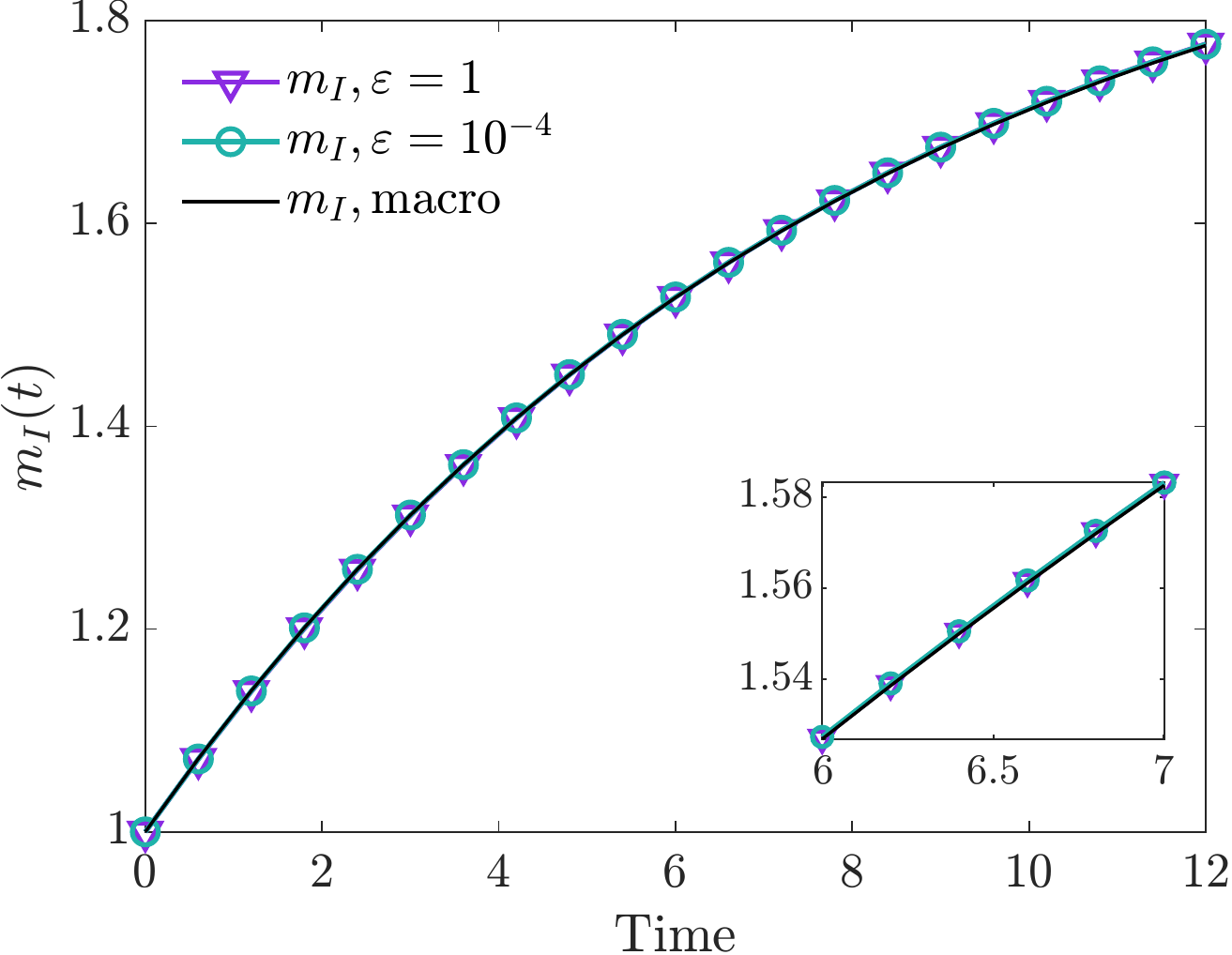}\\
\includegraphics[scale = 0.4]{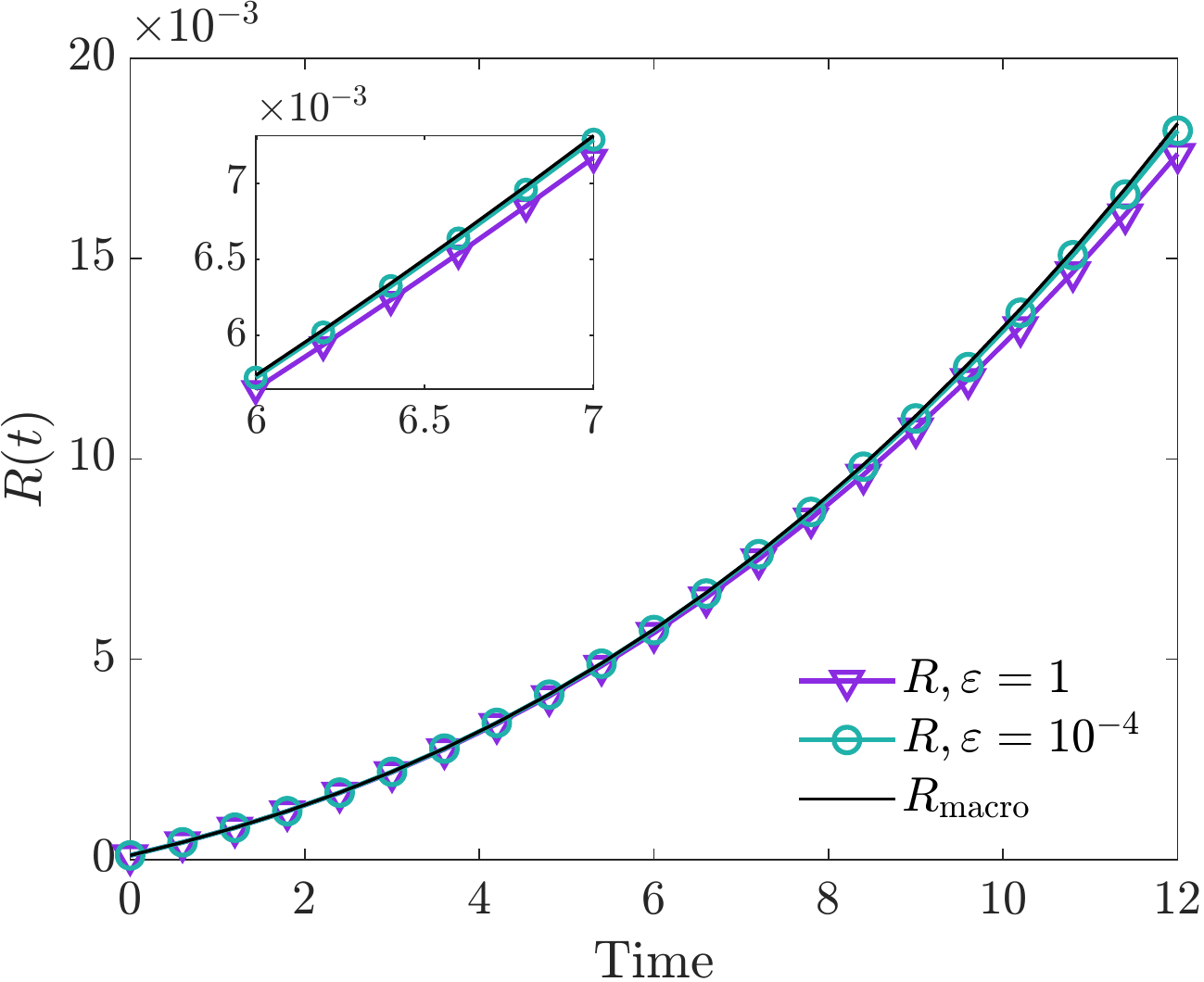}%
\includegraphics[scale = 0.4]{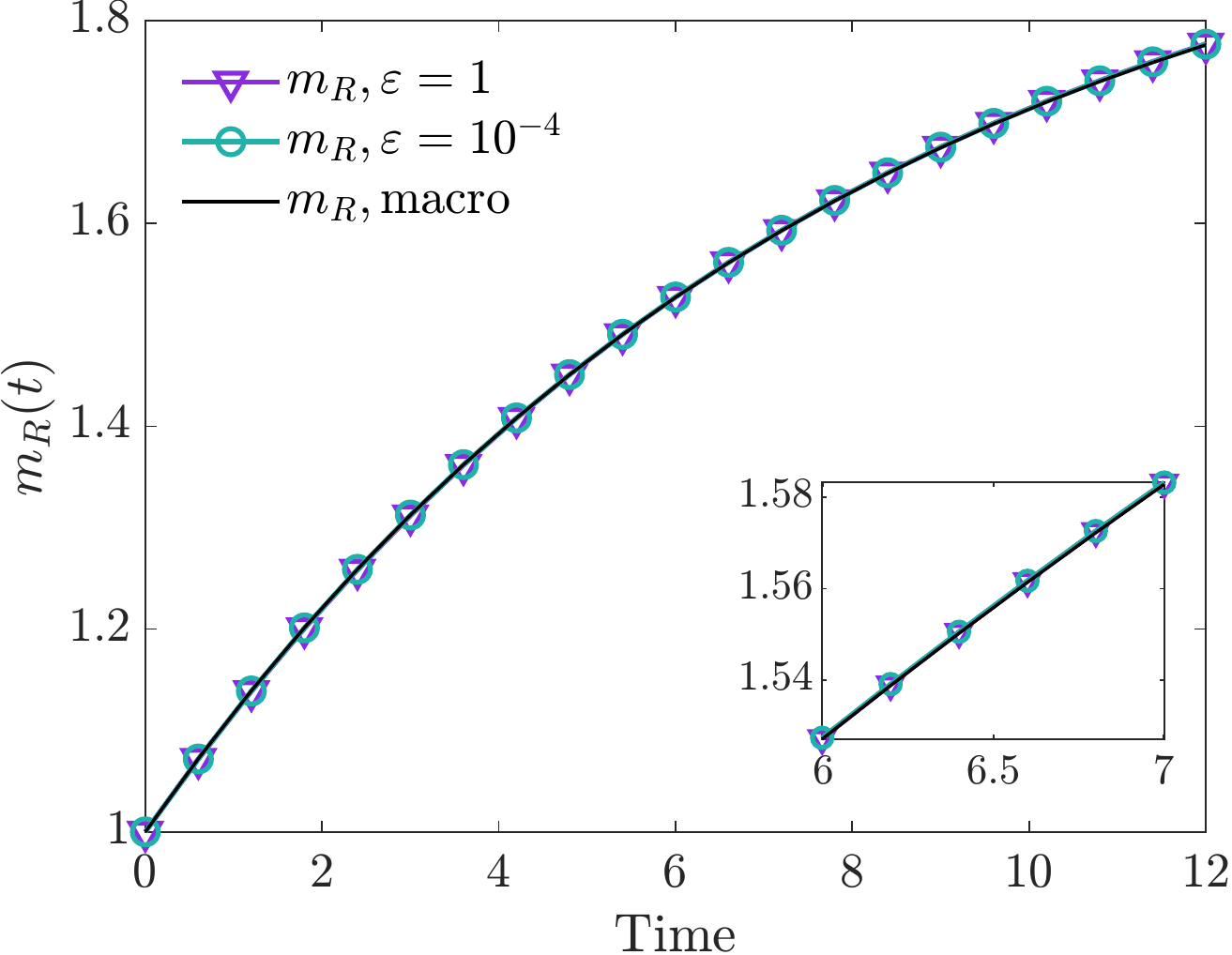}
\caption{\textbf{Test 1}. Evolution of mass fractions obtained from direct integration of the kinetic model \eqref{eq:seisc1} in the case $\kappa(x,x_*) = 
\beta e^{-x-x_*}$, for $\epsilon = 1$, 
$\epsilon = 10^{-4}$ together with the evolution of mass fractions of the 
macroscopic model \eqref{eq:seirAfr}--\eqref{eq:seirAme2}. In both cases we 
considered $\alpha = 0.9$, $\beta = 20$, $\gamma = 0.2$, $\delta = 0.05$. The 
kinetic model has been solved through the scheme $\mathrm I$--$\mathrm{II}$ 
over the domain $[0,4]$, discretization obtained with $N_v = 201$ grid points 
and $\Delta t = 10^{-4}$. The initial distribution of the kinetic model has 
been defined in \eqref{eq:f0_t1}-\eqref{eq:m0_t1}. }
\label{fig:epsilon-exp}
\end{figure}

In the test of this subsection, unless otherwise specified, we prescribe as initial datum the distribution
\begin{equation}
\label{eq:f0_t1}
f_H(x,0) = H(t) \dfrac{a_1^{a_2}}{\Gamma(a_2)}x^{-a_2-1}\exp\{-a_1/x\}, \qquad H \in \{S,E,I,R\},
\end{equation}
where $a_1 = 2(a_2-1)$ and $a_2 = 1.25$ with initial mass fractions 
\begin{equation}
\label{eq:m0_t1}
S(0) = 0.98, \qquad E(0) = 0.018, \qquad I(0) = R(0) = 0.001. 
\end{equation}
We consider the choice of parameters $m = \sum_{H \in 
\mathcal C}H(0)m_H(0)$, $\lambda = 0.25$ and $\sigma = 0.01$ for 
\eqref{eq:FP1}. The fake news dynamics is regulated by the following choice of 
parameters $\alpha = 0.9$, $\beta = 20$, $\gamma = 0.2$, and $\delta = 0.05$. 
For contact rates \ref{item:strong}--\ref{item:weak} we compared the 
evolution of mass fractions and mean values obtained from the integration of 
\eqref{eq:seisc1} with the ones derived in Section \ref{sec:red}. We 
consider the time interval  $[0,T]$, $T = 12$, a uniform time discretization 
with $\Delta t = 10^{-4}$ and $\epsilon = 1$,^^>$10^{-4}$. In particular, 
Figure^^>\ref{fig:epsilon-ip} refers to the case $\kappa(x,x_*)  = \beta/xx_*$ 
and Figure \ref{fig:epsilon-exp} to the case $\kappa(x,x_*)= \beta e^{-x-x_*}$. 
In both cases we may observe that for small values of $\epsilon$ the obtained 
macroscopic models are accurate in describing the trends of observable 
quantities of the kinetic field model. The macroscopic systems of coupled ODEs 
has been solved through a RK4 numerical scheme with $\Delta t = 10^{-4}$. 

\subsection{Test 2: A data driven application to Twitter}

In this test we focus on the spreading of the fake news by considering available Twitter data from the repository TweetSets\footnote{Justin 
Littman. (2018). TweetSets. Zenodo. 
\url{https://doi.org/10.5281/zenodo.1289426}}. In details, we 
analyzed the evolution from March to November, 2020 of the hashtag 
\texttt{\#facemask} related to the COVID-19 pandemic, and of the hashtags 
\texttt{\#hurricaneflorence\#fakenews} both associated to the hurricane 
Florence of September 2018 that caused catastrophic damages in USA, 
particularly in the states of North Carolina and South Carolina. 

\begin{figure}
	\includegraphics[width=0.45\linewidth]{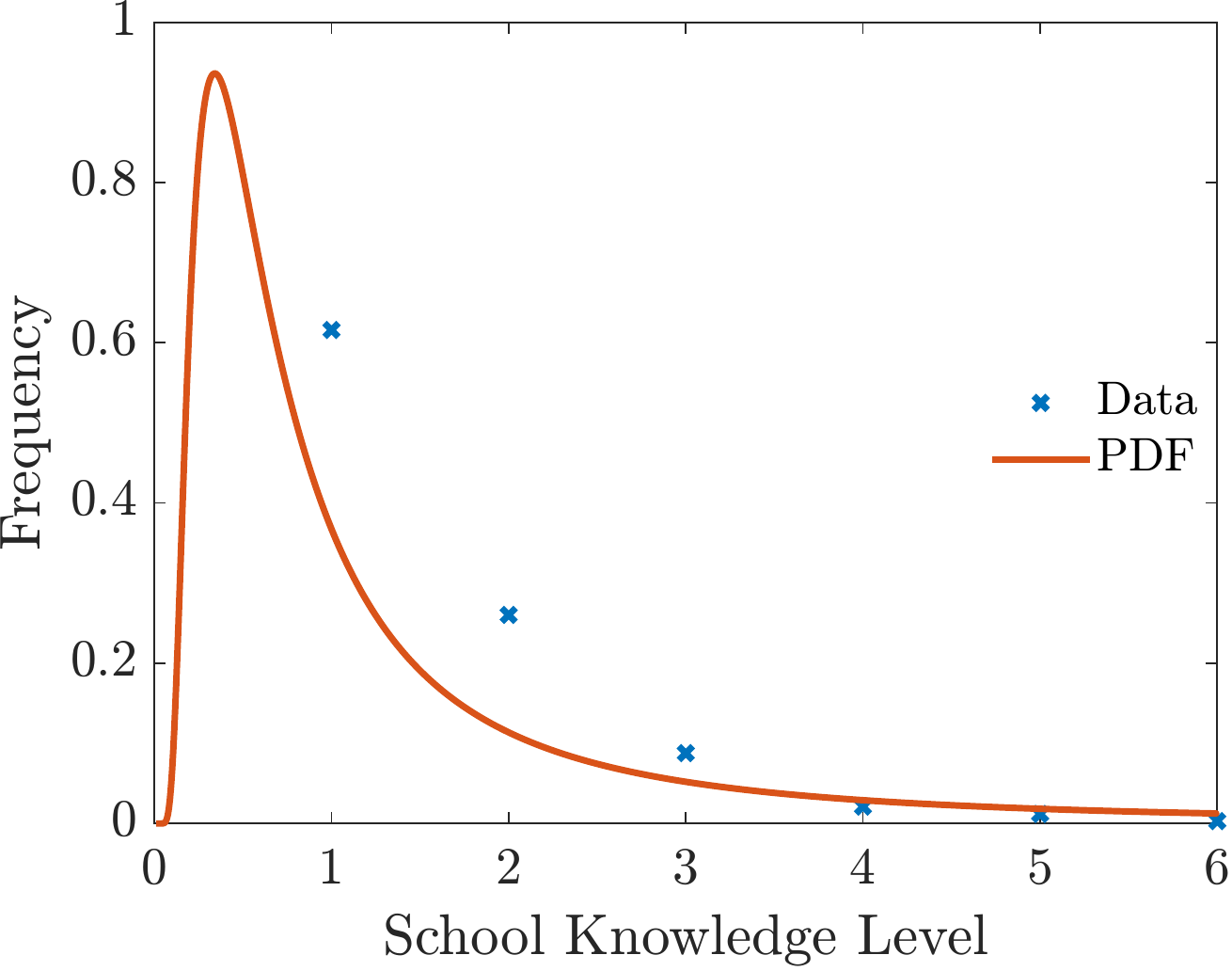}\hskip .4cm
	\includegraphics[width=0.45\linewidth]{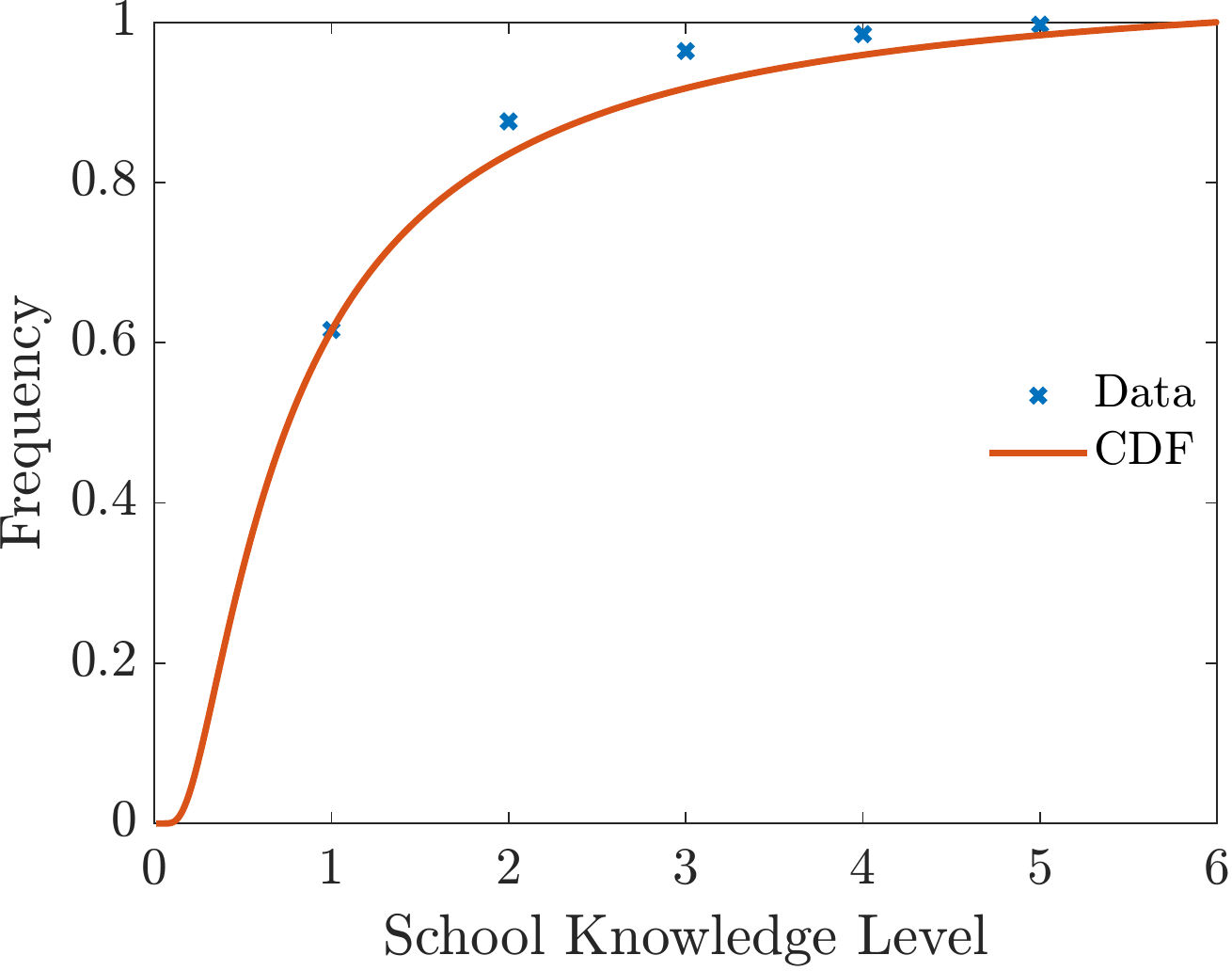}
	\caption{\textbf{Test 2}. Competence distribution and its inverse Gamma approximation $f(x)$ \eqref{eq:co_t2} corresponding to $c_1 \approx 0.75$, $c_2 \approx 1.25$ and leading to a mean competence background of $m_B = 3$. Data refers to 2011 Italian census and are used as representative of a prototypical Western country. On the $x$-axis we indicated with  $(1)$ lower secondary education, $(2)$ upper secondary education, $(3)$ undergraduate, $(4)$ master, $(5)$ second level master, $(6)$ doctorate. }
	\label{fig:gammafun}
\end{figure}

In the following we will assume that the competence variable is strongly related to the education level of a country. The data for the initial distribution of education has been extrapolated by the available Italian data from 2011 ISTAT census, and has been considered as representative data of a prototypical Western country^^>\cite{ToscaniGualandi}. As underlined in^^>\cite{ToscaniGualandi} the cumulative distribution of education exhibits a power-law type of tail. For this reason, as an approximation of the competence distribution we considered an 
inverse Gamma of the form
\begin{equation}
\label{eq:co_t2}
g(x) = \dfrac{c_1^{c_2}}{\Gamma(c_2)}\dfrac{e^{-c_1/x}}{x^{1+c_2}},  
\end{equation}
with $c_1,c_2>0$ obtained by data fitting. 
More precisely, we measure the education level on the scale $[0,6]$ where 6 represents the education of people with a PhD (see Figure \ref{fig:gammafun}). 

\subsubsection{Test 2A: Fitting the model to data}

Once we have obtained the initial competence distribution together with the 
value of $m_B$ we can estimate the parameters of the models defined in 
\eqref{eq:seirAfr}--\eqref{eq:seirAme2} and 
\eqref{eq:seirBfr2}--\eqref{eq:seirBme2}. Several approaches have been proposed 
in the literature, see e.g., \cite{liu2020}. It is worth to mention that 
several uncertainties are present in data linked to news-monitoring. For 
example the total population size is generally unknown and the total number of 
Twitter accounts represent an upper bound over the real active users. 

The approach adopted in^^>\cite{Fang2013}, and subsequently 
in^^>\cite{Fang2014,Maleki2021}, is to treat this quantity as a parameter to be determined in the minimization process along with the parameters 
of the models. To reduce the number of parameters to optimize we follow a different path. In particular, as initial guess on the total population size, since the datasets that we 
used for the fitting were based on U.S. hashtags, we considered that each 
fake-news spreader has in average 453 followers^^>\footnote{Kickfactory, link at 
\url{https://kickfactory.com/blog/average-twitter-followers-updated-2016/}, 
last accessed: 21th February 2022.}. Hence, in average we may expect that the 
total number of susceptible is given by the total number of tweets multiplied 
by the average number of followers. To take also into account both the number 
of bots 
on Twitter as found in^^>\cite{uyheng2020} (and references therein) and users 
whose activity could be not assiduous enough to matter during the lifespan of 
the considered fake news, the initial guess was also reduced by a factor of^^>$4$.

Let us denote by $\hat I(t)$ the number of active spreaders obtained from the 
data, while $I(t)$ is the number of infectious agents given by the macroscopic 
differential model. Hence, we consider the following cost functional 
\[
F(\hat I, I) = \normap[\bigg]{ \int_{t_0}^{t_f} \hat I(t)dt - \int_{t_0}^{t_f} 
I(t)dt}{L^2([t_0,t_f])},
\]
where $[t_0,t_f]$ is the time-frame (in hours) during 
which we solve the minimization problem
\begin{equation}\label{eq:minproblem}
\min_{\alpha,\beta,\gamma,\delta \in \mathbb R_+} F(\hat I, I),
\end{equation}
whereas $\eta$ was kept fixed and equal to $0.5$. 

Since data for the evolution of compartments $S,E,R$ are not at our disposal, as well 
is not the initial means value for any of the compartments, we solved the ODE 
model on $[t_\star, t_0]$, where $t_0$ is the starting point of the spreading 
process and $t_\star$ is a suitable unknown time previous to $t_0$ starting from single   exposed, infectious and recovered individuals. The idea is 
to simulate an initial situation for the spread of fake news to happen. Furthermore, we considered initial mean values equal to the half of the mean background distribution of competence, i.e. $m_J(0) = 1.5$.

\begin{figure}
\centering
\includegraphics[width=0.325\linewidth]{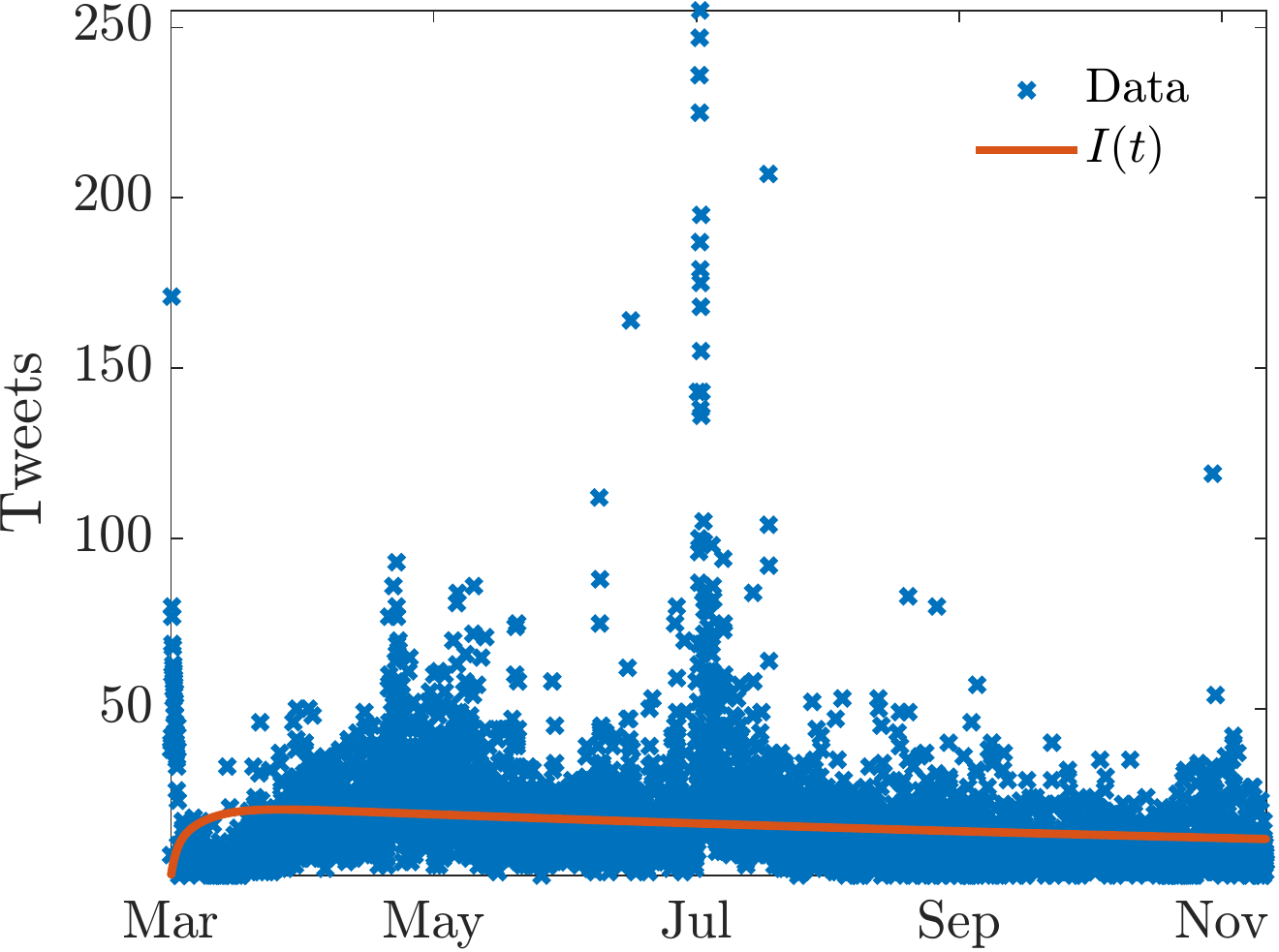}%
\includegraphics[width=0.325\linewidth]{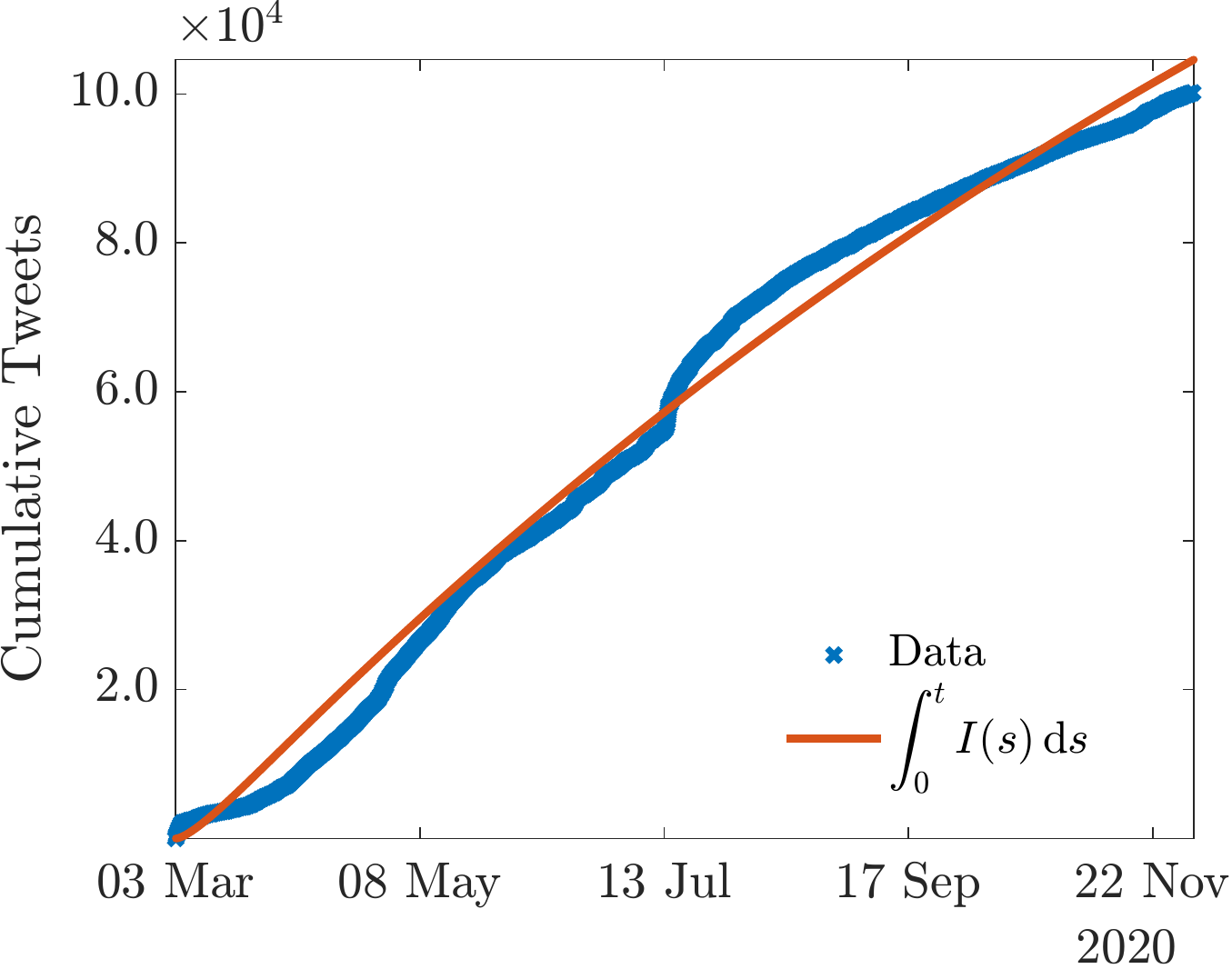}%
\includegraphics[width=0.325\linewidth]{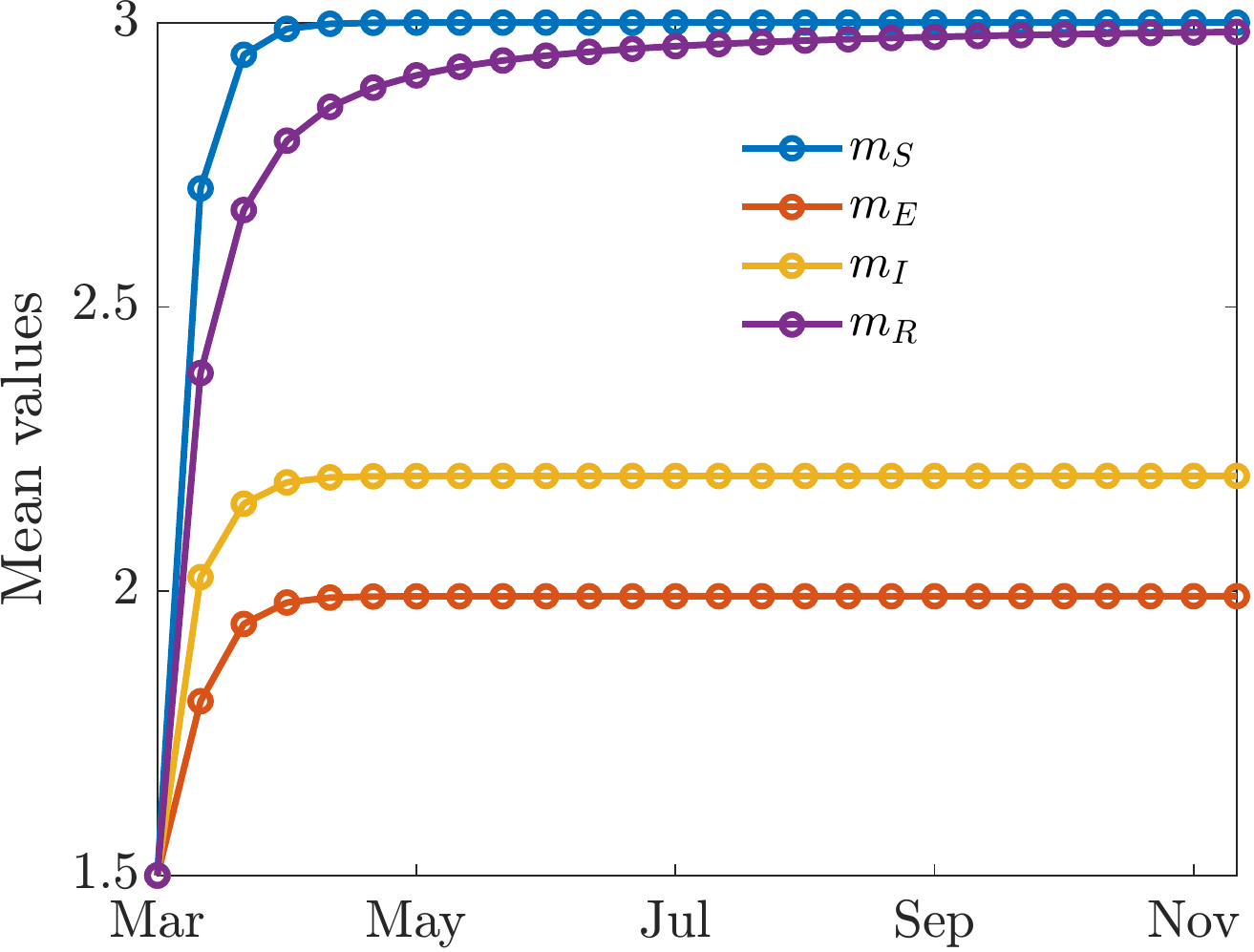}\\
\includegraphics[width=0.325\linewidth]{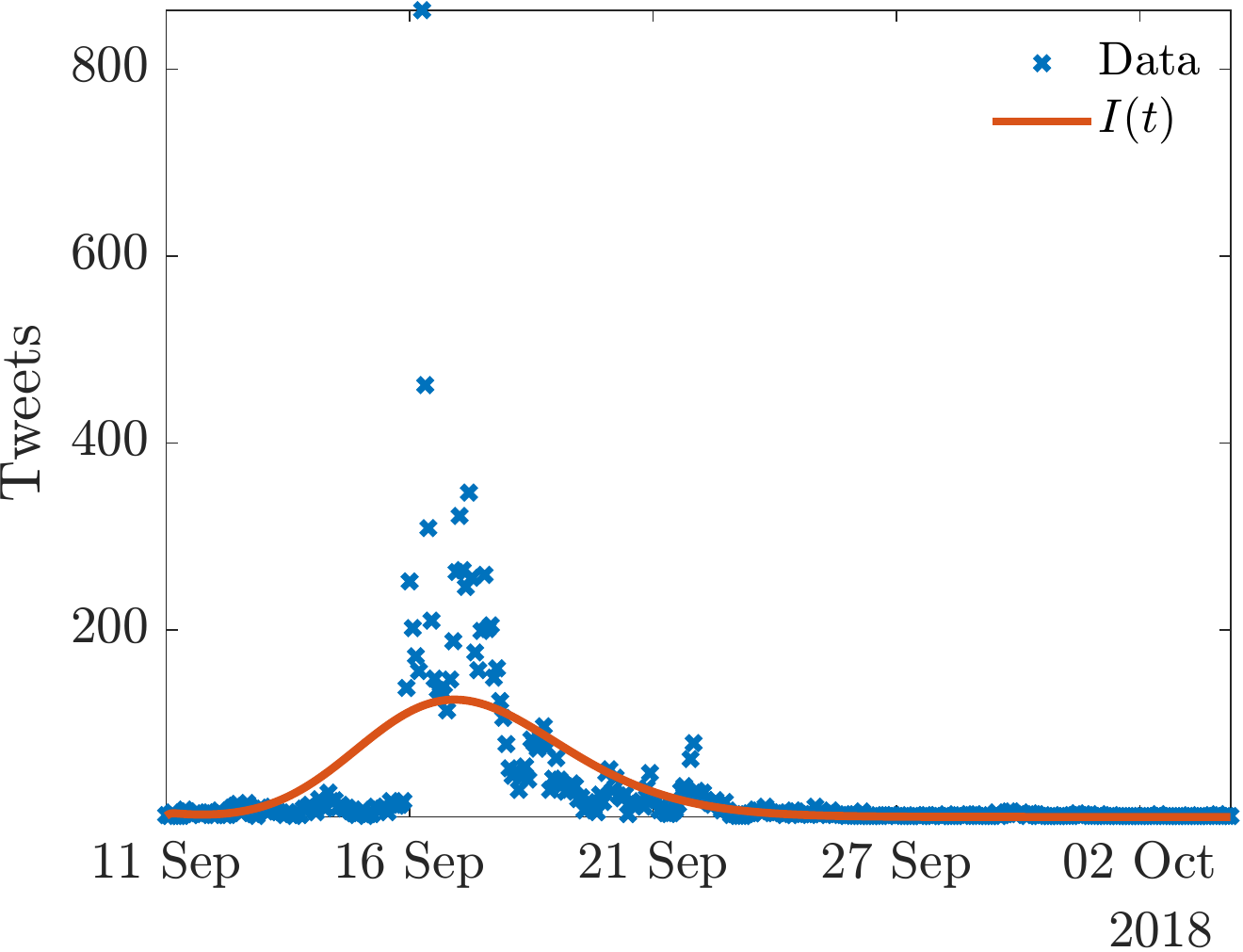}%
\includegraphics[width=0.325\linewidth]{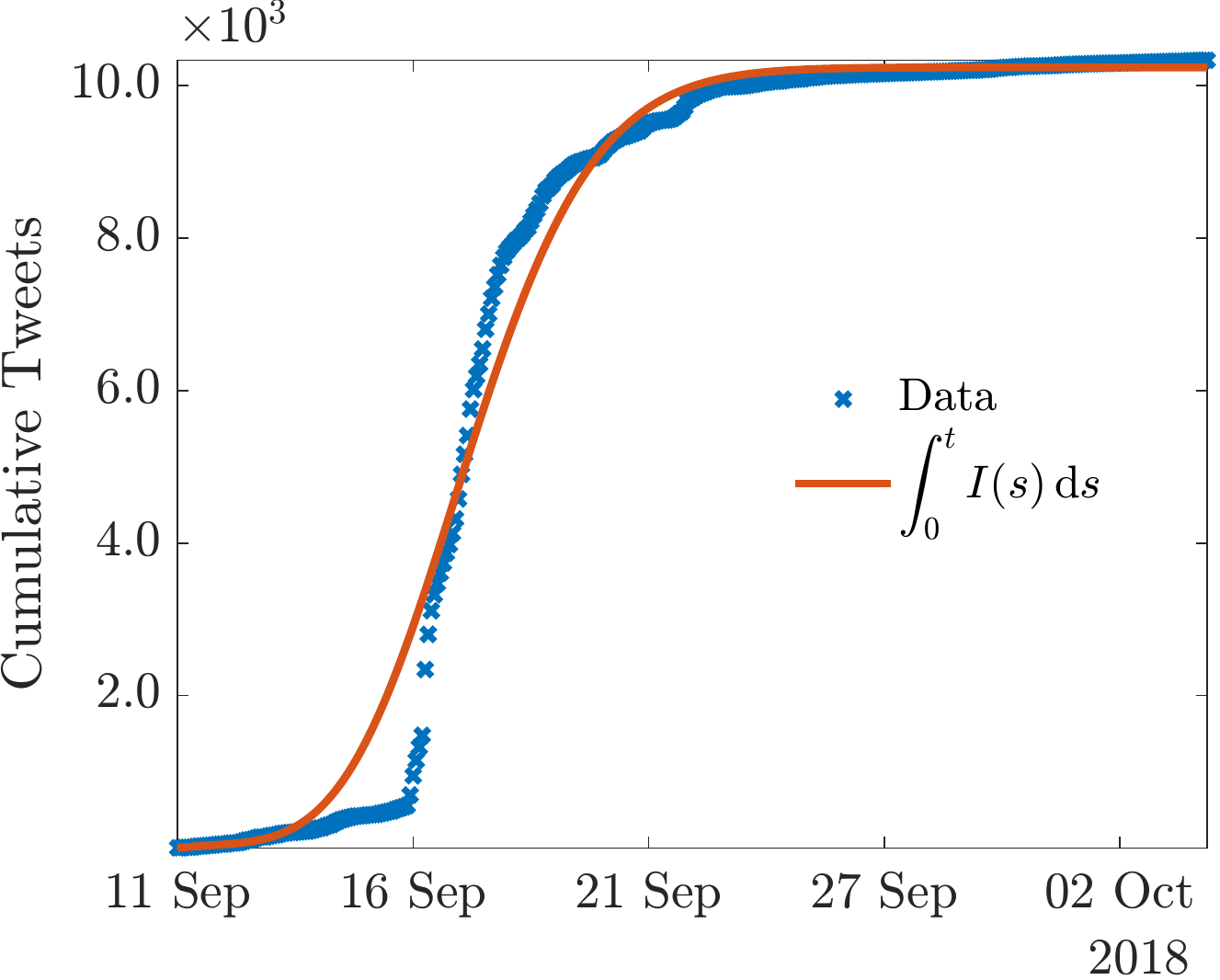}%
\includegraphics[width=0.325\linewidth]{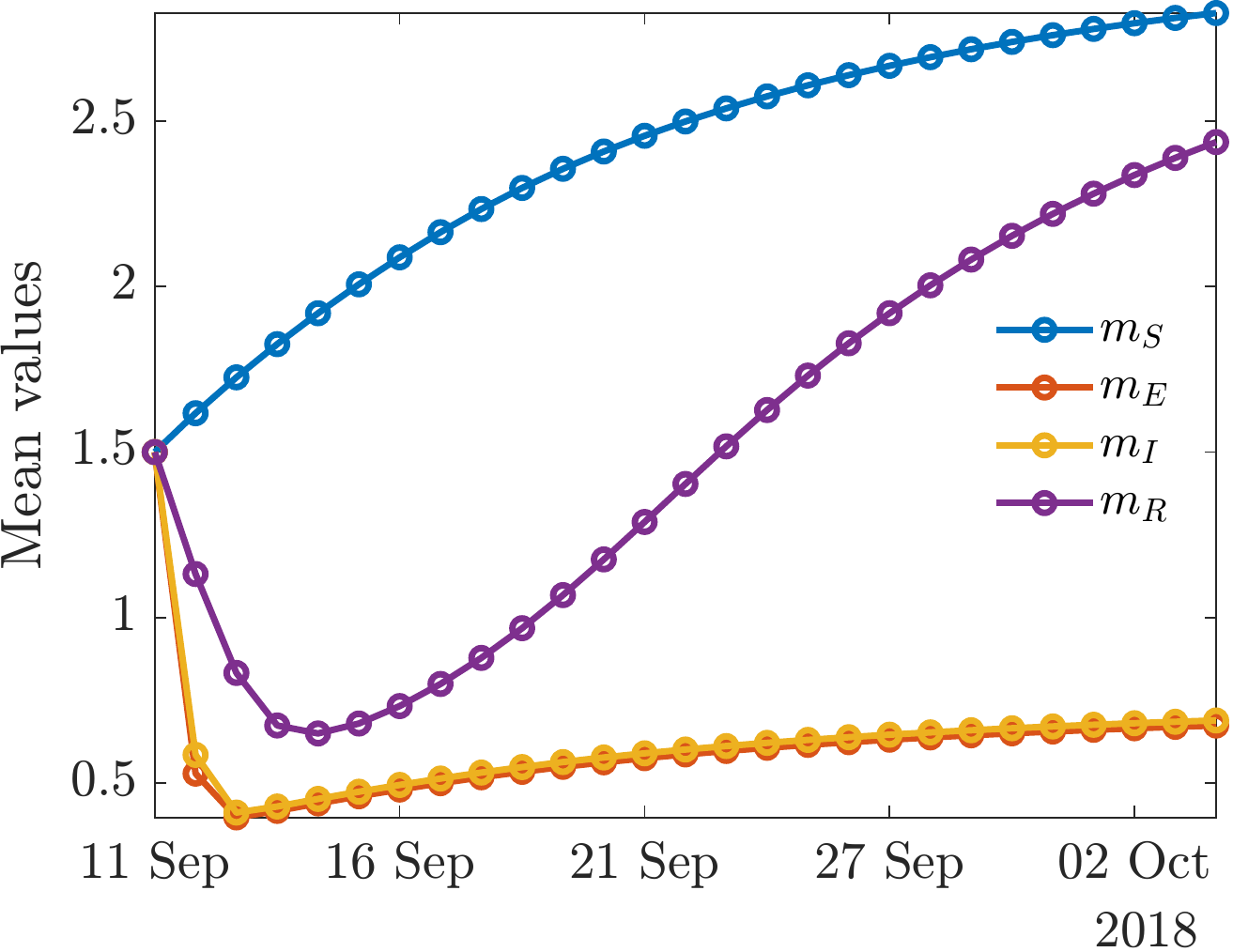}
\caption{\textbf{Test 2A.} Top row: optimization results for 
\texttt{\#facemask}. Bottom row: optimization results for  
\texttt{\#florence\#fakenews}. From right to left: approximation of raw data on 
the number of tweets, cumulative distribution and evolution of the mean 
competence for each compartment. }\label{fig:facemaskconstant}
\end{figure}

\begin{table}
\centering
\begin{tabular}{ccccc}
\toprule
Parameter & \multicolumn{2}{c}{\texttt{\#facemask}}
          & \multicolumn{2}{c}{\texttt{\#florence\#fakenews}}\\ 
\cmidrule{2-5}
          &$\kappa(x,x_*) = \beta/(x\, x_*)$ & $\kappa(x,x_*)=\beta e^{-x-x_*}$
          &$\kappa(x,x_*) = \beta/(x\, x_*)$ &$\kappa(x,x_*)=\beta e^{-x-x_*}$\\
\midrule
$\alpha$  & 0.9995  & 0.9993   & 1.0000  & 0.9999 \\
$\beta$   & 0.0122  & 0.2937   & 0.0901  & 0.9999\\
$\delta$  & 0.0237  & 0.0336   & 1.0000  & 0.1930\\
$\gamma$  & 0.0046  & 0.0079   & 0.2127  & 0.9999\\
\bottomrule
\end{tabular}
\caption{\textbf{Test 2A.} Estimated parameters for the entire datasets for the 
hashtags \texttt{\#facemask} (second and third column) and 
\texttt{\#florence\#fakenews} (fourth and fifth column).}\label{tab:test2params}
\end{table}

In Figure^^>\ref{fig:facemaskconstant} we compare the evolution on the number 
of tweets regarding the hashtag \texttt{\#facemask}, from 3rd March 2020 to 22 
November 2020, and the hashtag  \texttt{\#florence\#fakenews}, from 11th 
September 2018 to 4th October 2018, with the evolution of the model 
\eqref{eq:seirAfr}, \eqref{eq:seirAme2}. The obtained parameters are reported 
in Table^^>\ref{tab:test2params}.

In both cases,  we may observe that the evolution of the mean competence levels are different in the four compartments and, in particular, that low competence levels are associated to exposed and infectious agents, i.e., the 
active spreaders. The outcome reflects the intuitive idea that the disinformation could be 
driven by the lack of capability to recognize an information as purposely false 
in the first place. 

To better take into account the impact of a competence-based contact rate 
function $\kappa(x,x_*)$, we also computed the associated basic reproduction 
number $R_t$ using  the parameters $\beta$ 
and $\gamma$ estimated previously for both datasets, reported in 
Table^^>\ref{tab:test2params}.
Following^^>\cite{bertaglia2021,franceschi2022}, and omitting the details for 
brevity, we consider a generalized version of the classical reproduction number 
defined as
\begin{equation}\label{eq:reproductionnumber}
R(t) = \frac{\int_{\R^+} K(f_S,f_I)(x,t)\,dx}{\int_{\R^+} \gamma f_I(x,t)\,dx},
\end{equation}
where again we leveraged the structure preserving scheme proposed 
in^^>\cite{pareschi2018} to perform the calculations.
\begin{figure}
\centering
\includegraphics[width=0.45\linewidth]{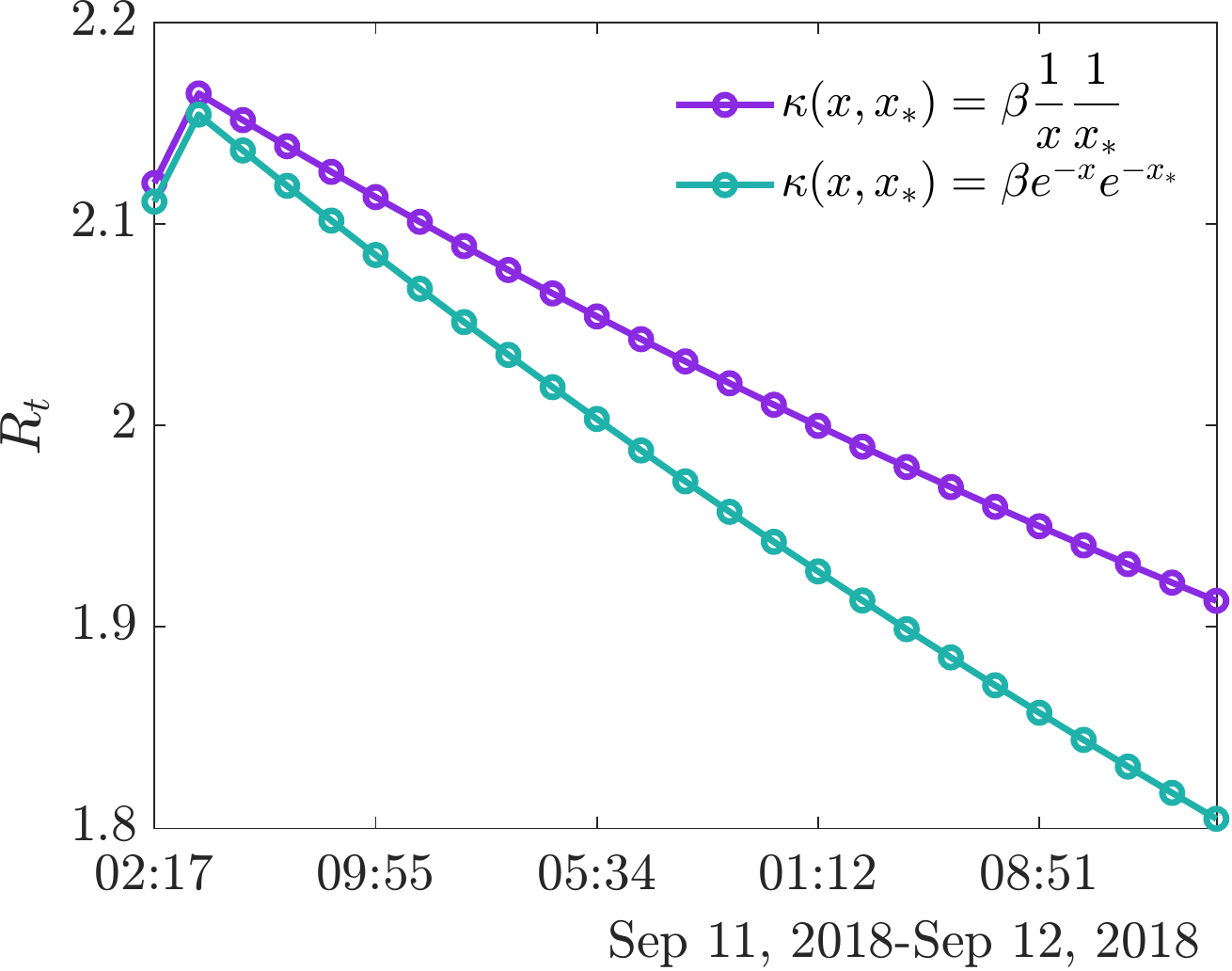}^^>%
\includegraphics[width=0.45\linewidth]{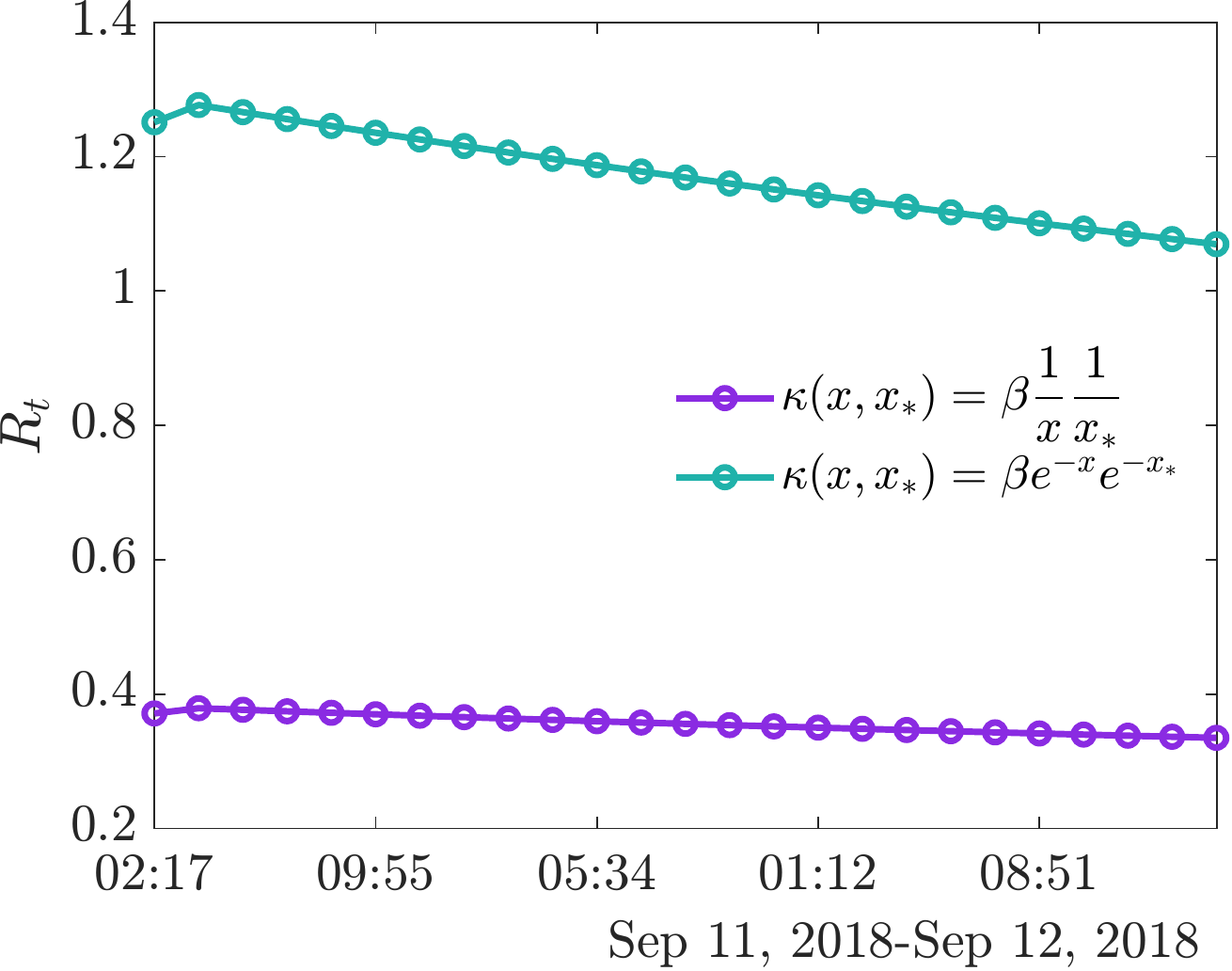}^^>%
\caption{\textbf{Test 2A}. Evolution of $R_t$ in the first 24 hours of datasets 
\texttt{\#facemask} (left) and \texttt{\#florence\#fakenews} (right) for the 
parameters estimated in Table^^>\ref{tab:test2params} relative to 
the introduced contact functions $\kappa(x,x_*)$. }
\label{fig:R0}
\end{figure}

\subsubsection{Test 2B: Forecasting under data uncertainties}
To analyze the impact of uncertainties in data and parameters we consider a 3D random variable $\mathbf z = (z_1,z_2,z_3)$ with distribution $\rho(\mathbf z)$. We will suppose that the random vector $\mathbf z$ has independent components, i.e. $\rho(\mathbf z) = \rho_1(z_1)\rho_2(z_2)\rho_3(z_3)$. Taking into account parametric uncertainties, we consider the estimated model parameters as follows
\begin{equation}
\label{eq:stoc}
\beta(z_1) = \beta_0 (1+ c_\beta z_1),\qquad \gamma(z_2) = \gamma_0 (1+ c_\gamma z_2), \qquad \delta(z_3) = \beta_0 (1+ c_\delta z_3), 
\end{equation}
where we supposed $z_1,z_2,z_3 \sim \mathcal U([-1,1])$ and $c_\beta,c_\gamma,c_\delta>0$. As a result, the macroscopic quantities describing the evolution of compartments result affected by the introduced uncertainties increasing their dimensionality $J(\mathbf z,t)$, $m_J(\mathbf z,t)$, $J \in \mathcal C$. 
 In order to handle efficiently the introduced uncertainties in the dynamics we adopt a stochastic collocation approach based on stochastic Galerkin methods, we refer the interested reader to^^>\cite{Xiu} for an introduction and to^^>\cite{APZ1,zanella_2021} for applications in compartmental modelling of epidemic dynamics. This class of methods allows to accurately quantify the propagation stochasticity in a parametric differential model when information on the uncertainties' distribution are available. We remark that fast convergence properties hold under suitable regularity assumptions on the problem's solution. In details, we construct a 3D sample $\{z_{i,k}\}_{k=0}^M$, $i=1,2,3$, obtained in a collocation setting through Gauss-Legendre polynomials with $M = 5$ nodes.

\begin{figure}
\centering
\includegraphics[scale = 0.6]{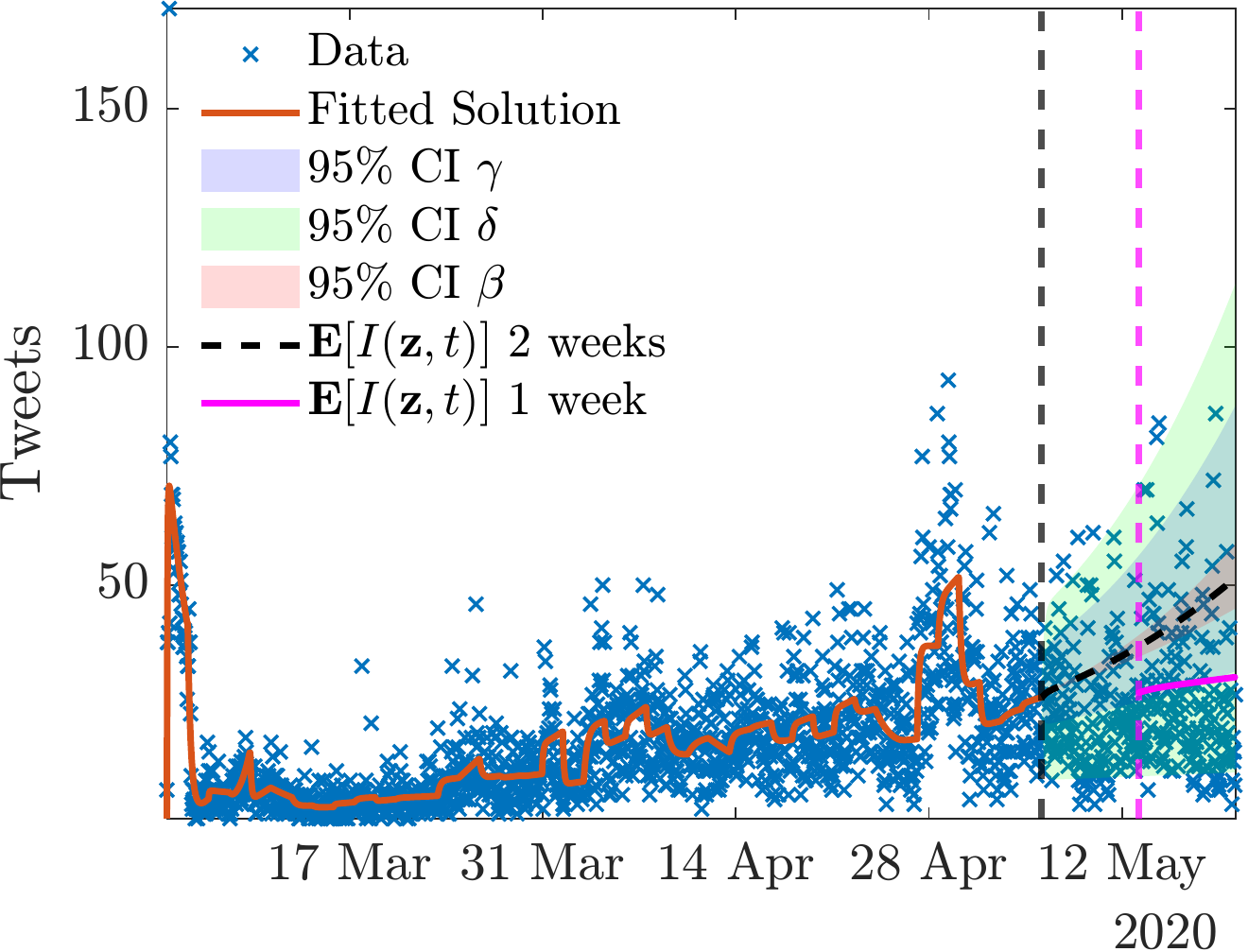}\hskip .4cm
\includegraphics[scale = 0.6]{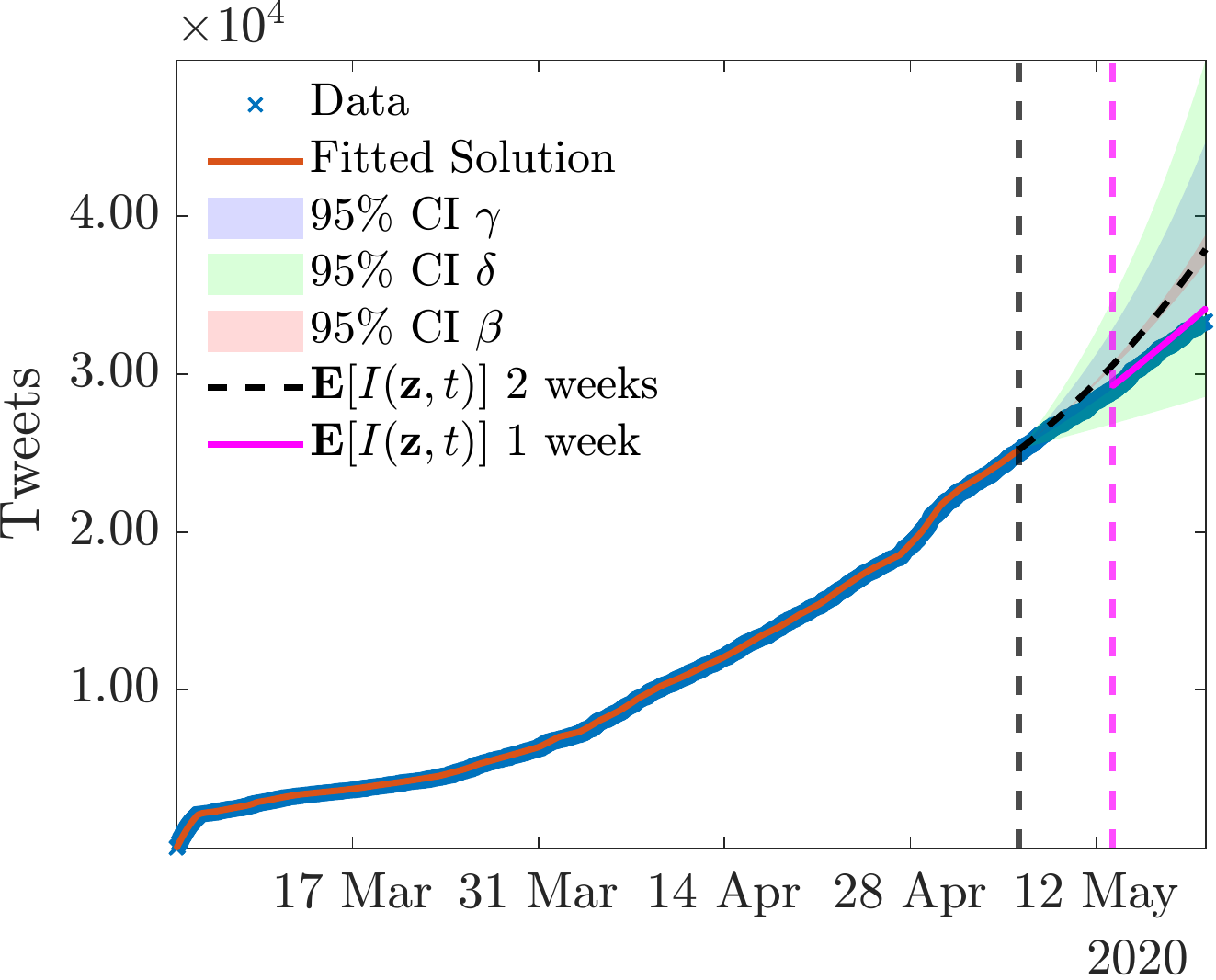} \\ 
\includegraphics[scale = 0.6]{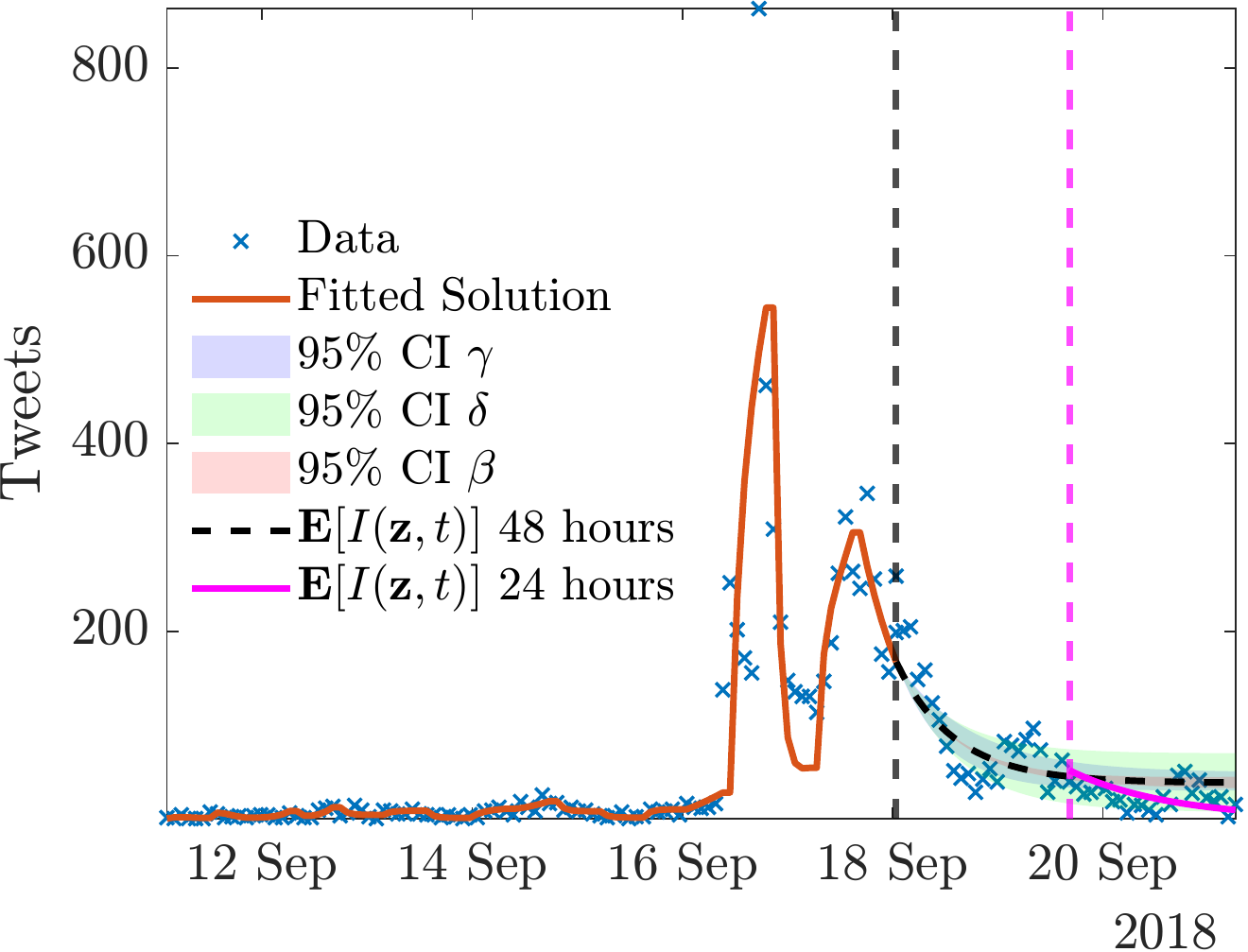} 
\includegraphics[scale = 0.6]{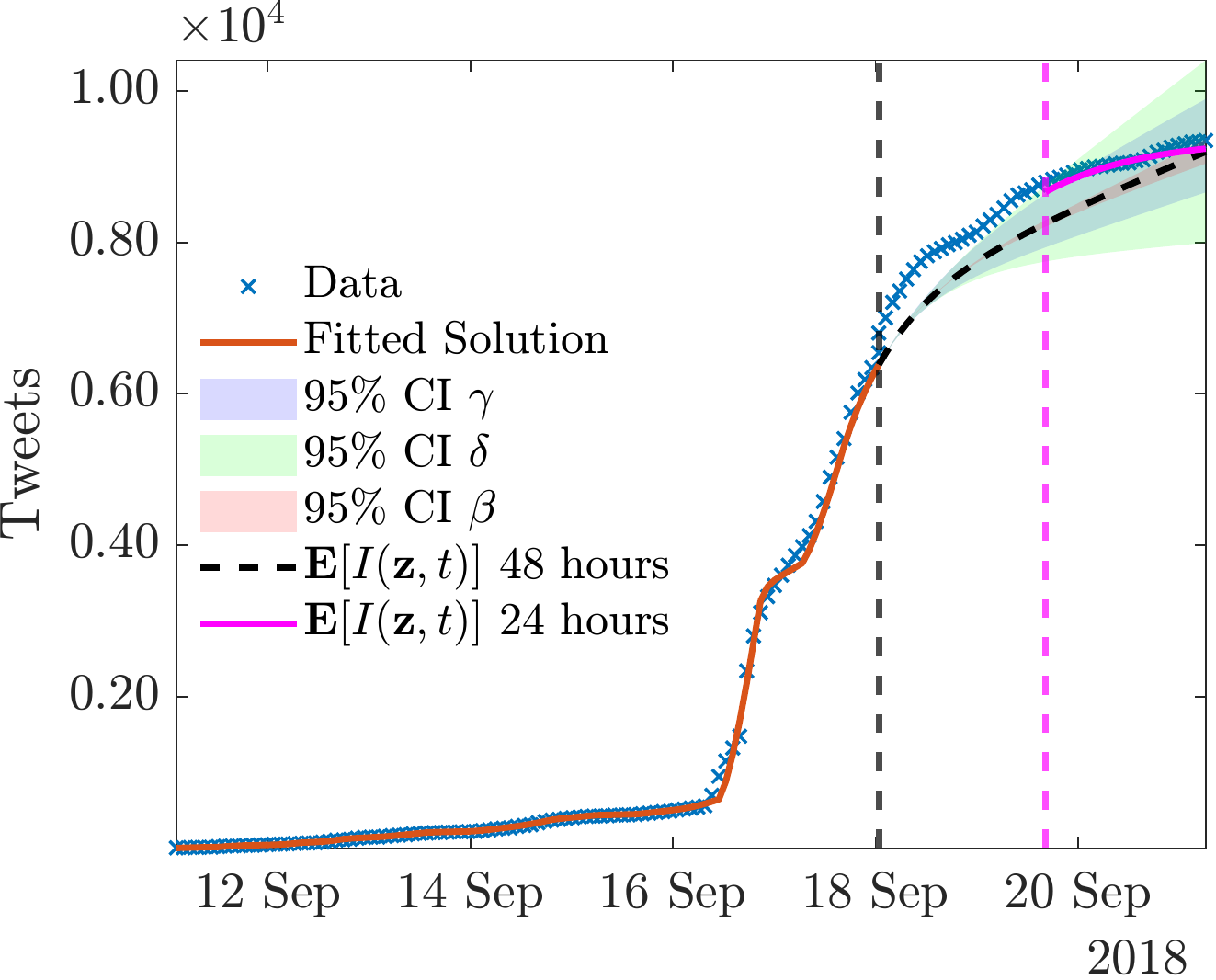}
\caption{\textbf{Test 2B}. Top row: comparison between 14 days (May 3rd--May 14th, 2020) and 7 days (May 3rd--May 10th, 2020) predictions of \texttt{\#facemask} based on the model \eqref{eq:seirAfr}, \eqref{eq:seirAme2} with uncertain parameter \eqref{eq:stoc} with $c_\beta = 200 \beta_0$, $c_\gamma = 2 \gamma_0$, $c_\delta = 2\times 10^4 \delta_0$. Bottom row: comparison between 24 hours (September 18th-September 19th, 2018) and 48 hours (September 18th-September 20th, 2018) predictions of  \texttt{\#florence\#fakenews} based on the model  \eqref{eq:seirAfr}, \eqref{eq:seirAme2} with uncertain parameter \eqref{eq:stoc} with $c_\beta = 0.2 \beta_0$, $c_\gamma =  \gamma_0$, $c_\delta = 2\times 10^4 \delta_0$. }
\label{fig:UQ}
\end{figure}

In Figure^^>\ref{fig:UQ} we display the dynamics of the considered fake-news with respect to available data. In details, for \texttt{\#florence\#fakenews} we consider the period  from September 11th 2018 to September 21st 2018. We consider two successive prediction horizons respectively of 1 day, i.e. the parameters of the models are calibrated taking into account data until September 19th,  and a 2 days prediction horizon, where the calibration is based only on data until September 18th. 
Regarding \texttt{\#facemask} we considered the period from March 3rd to May 17th. Also in this case we consider  two successive prediction horizons of 1 week,  i.e. the parameters of the models are calibrated taking into account data until May3rd, and a two weeks prediction horizon, where the calibration is based on data until May10th. 

We highlight in dashed black and magenta the expected value of the predicted number of tweets $\mathbb E[I(\mathbf z,t)] = \int_0^1 I(\mathbf z,t)\rho_1(z_1)\rho_2(z_2)\rho_3(z_3)dz_1 dz_2 dz_3$. Together with the expected trends we plot the $95\%$ confidence intervals (CI) with respect to the random parameters $\beta(z_1)$, $\gamma(z_2)$, and $\delta(z_3)$. The blue shaded band is relative to the variability in $\gamma(z_2)$, the green shaded to the variability in $\delta(z_3)$ whereas the shaded red is relative to the variability in $\beta(z_1)$.

\subsection{Test 3: Competence background in misinformation}

In this test we perform a retrospective analysis to study how the background could influence the dissemination of fake news as a result of a different learning process. We recall that the background modifies through a learning dynamic the effectiveness of the level of knowledge in identifying fake news. As a consequence high values of the background correspond to a high level of effectiveness of the competence while low values will make it difficult to identify the fake-news. 
Indirectly, the background acts as a control term which limits the spread of the misinformation. This can also be interpreted as a process of education specific to the identification of fake news that allows to limit the so-called knowledge neglect phenomenon^^>\cite{Fazio}.  

We consider the two datasets for the 
hashtags \texttt{\#facemask} and \texttt{\#florence\#fakenews} with the 
estimated parameters reported in Table^^>\ref{tab:test2params} and we  
increase the value of the competence level attained by the background, i.e., 
$m_B$, while keeping fixed the parameters during the dynamics defined by \eqref{eq:seirAfr},\eqref{eq:seirAme2} and 
\eqref{eq:seirBfr2},\eqref{eq:seirBme2}. 

\begin{figure}
\centering
\includegraphics[width=0.45\linewidth]{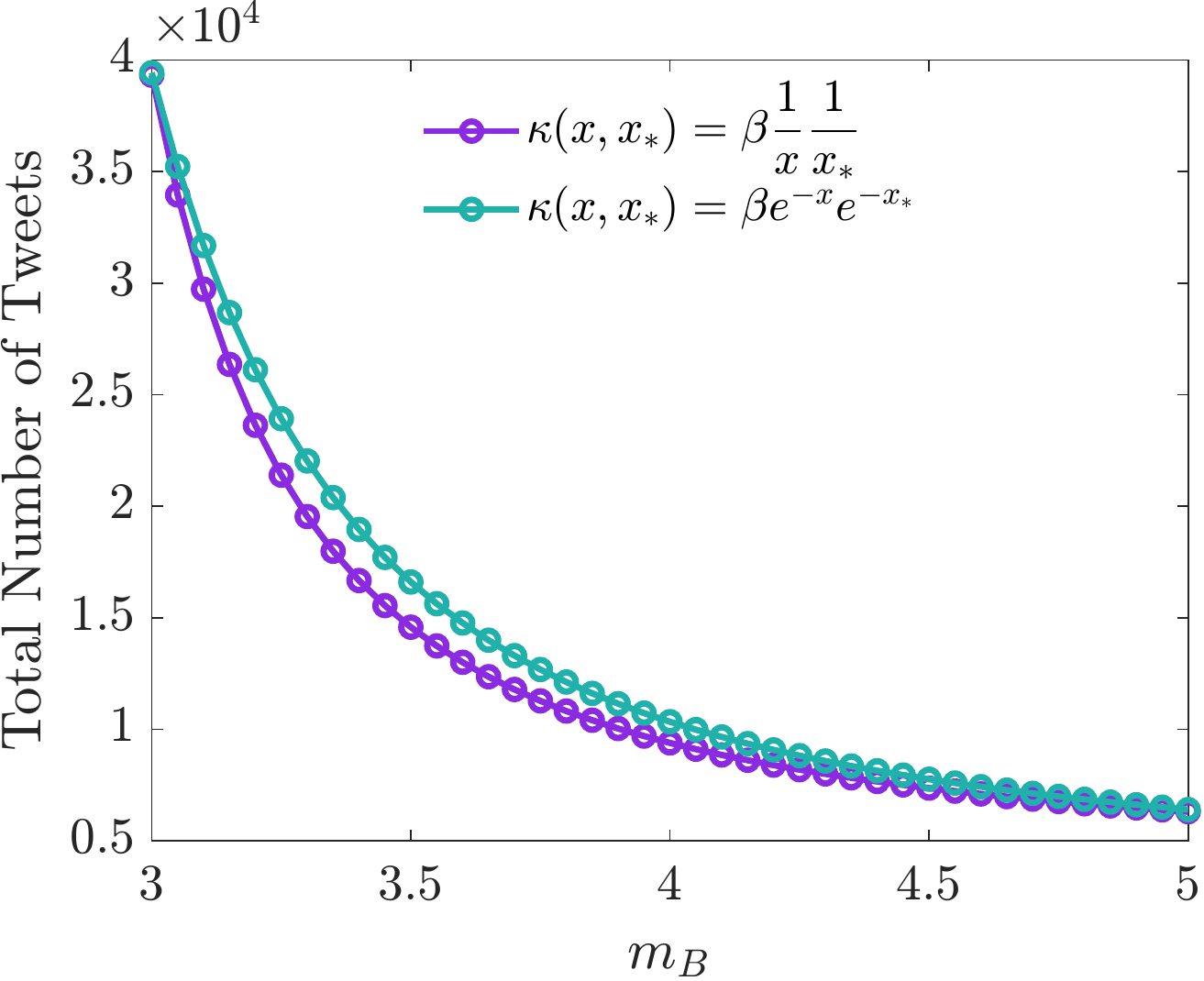}\hskip .5cm
\includegraphics[width=0.45\linewidth]{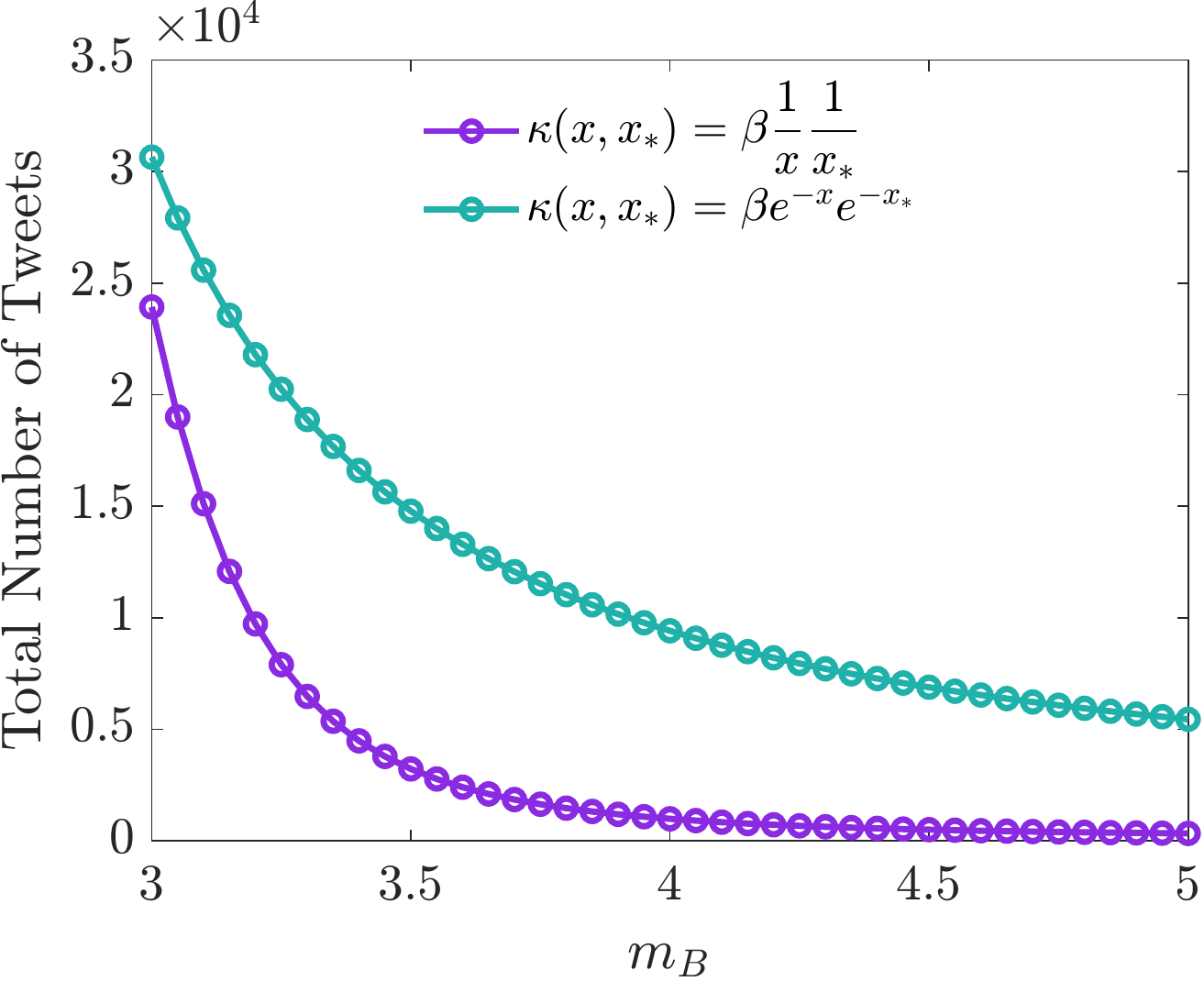}%
\caption{\textbf{Test 3.} Total number of infectious agents for the hashtags 
\texttt{\#facemask} (left) and \texttt{\#florence\#fakenews} (right) as a function 
of the competence background. 
In both cases, we employed 
the parameters reported in Table^^>\ref{tab:test2params}.}
\label{fig:test3}
\end{figure}

Hence, we performed the test with both choices of a strong and weak competence based 
contact function; the results are summarized in Figure^^>\ref{fig:test3}. 
In all cases, we see how increasing the competence of the background reduces the 
spread of fake news, leading to a decrease in the cumulative number of tweets of infectious agents proportional to the increase in the value of $m_B$. 
Indeed, we can observe how increasing the competence of the 
background, we obtain an evident decrease in the overall misinformation for both the 
examples considered \texttt{\#facemask} and \texttt{\#florence\#fakenews}.

\section*{Concluding remarks}
Despite the digital transformation of governments and the modernization of public administration, a global decline in democracy is occurring around the world. The spread of fake news created for the purpose of polarizing society in certain directions poses a risk to democratic institutions. The role of individuals' knowledge and the ability to use it in identifying false information is deemed of paramount importance. 

In this paper starting from a model for the description of fake-news dissemination
in the presence of heterogeneous agents with different levels of competence, through the tools of 
kinetic theory, reduced-order models have been derived that allow to keep the effects of the
of competence in the dynamics and that, thanks to their simplified structure, can be interfaced with data.

The starting model is inspired, as in much of the literature related to fake-news, 
to the epidemiology, so it is based on a compartmental structure. The introduction of competence
allows to analyze complex phenomena of great relevance in contemporary society,
such as the effectiveness of control actions taken to limit the spread of fake news and
the role of knowledge neglect in misinformation.

The methodology adopted in this article is fully general and depends closely on the equilibrium state of the social variable and the social interaction function at the basis of 
fake-news spreading. As a consequence, additional social variables that play
a key role in the spread of misinformation may be embedded in the dynamics using similar arguments. The ability to have a model that can be interfaced with the available data allowed us to present some preliminary examples of applications to the case of fake-news spreading on Twitter.

\section*{Competing interests}
The authors declare no competing interests. 

\section*{Data availability statement}
The datasets generated during the current study is available from the corresponding author on reasonable request. The datasets analysed during the current study are freely available from the website  \texttt{https://doi.org/10.5281/zenodo.1289426}.

\subsection*{Acknowledgments}
This work has been written within the activities of the GNFM and GNCS groups of 
INdAM (National Institute of High Mathematics). M.Z. acknowledge partial 
support of MUR-PRIN2020 Project \lq\lq Integrated mathematical approaches to 
socio-epidemiological dynamics\rq\rq. The research of M.Z. was partially 
supported by MIUR, Dipartimenti di Eccellenza Program (2018–2022), and 
Department of Mathematics “F. Casorati”, University of Pavia. The research of 
L.P. was partially supported by FIR project “No hesitation. For effective 
communication of Covid-19 vaccination”, University of Ferrara.

\phantomsection

\end{document}